\documentclass[11pt,a4paper]{article}
\pdfoutput=1
\usepackage{jcappub}
\usepackage[T1]{fontenc}
\usepackage[utf8]{inputenc}
\usepackage{amssymb}
\usepackage{soul}

\begin{document}
\hfill {\tt PI/UAN-2018-626FT}

\title{Sharing but not Caring:\\Collider Phenomenology}

\author[1]{Nicolás Bernal,}
\emailAdd{nicolas.bernal@uan.edu.co}

\author[2,3]{Chee Sheng Fong}
\emailAdd{cheesheng.fong@gmail.com}

\author[4,5]{and Alberto Tonero.}
\emailAdd{alberto.tonero@gmail.com}

\affiliation[1]{Centro de Investigaciones, Universidad Antonio Nariño\\
Cra 3 Este \# 47A-15, Bogotá, Colombia}

\affiliation[2]{Instituto de Física, Universidade de São Paulo\\
C.P. 66.318, 05315-970 São Paulo, Brazil}

\affiliation[3]{Departamento de F\'isica, Pontif\'icia Universidade Cat\'olica do Rio de Janeiro\\
Rua Marquês de São Vicente 225, Rio de Janeiro, Brazil}

\affiliation[4]{UNIFAL-MG\\
Rodovia José Aurélio Vilela 11999, 37715-400, Poços de Caldas, MG, Brazil}

\affiliation[5]{Ottawa-Carleton, Institute for Physics, Carleton University\\
1125 Colonel By Drive, Ottawa, ON, K1S 5B6, Canada}

\abstract{Based on a previous work on scenarios where the Standard Model and dark matter particles \emph{share} a common asymmetry through effective operators at early time in the Universe and later on decouple from each other (\emph{not care}), in this work, we study in detail the collider phenomenology of these scenarios. In particular, we use the experimental results from the Large Hadron Collider (LHC) to constrain the viable parameter space. Besides effective operators, we also constrain the parameter space of some representative ultraviolet complete models with experimental results from both the LHC and the Large Electron-Positron Collider. Specifically, we use measurements related to jets + missing transverse energy (MET), di-jets and photon + MET. In the case of ultraviolet models, depending on the assumptions on the couplings and masses of mediators, the derived constraints can become more or less stringent. We consider also the situation where one of the mediators has mass below 100~GeV, in this case we use the ultraviolet model to construct a new effective operator responsible for the sharing of the asymmetry and study its phenomenology.
}

\maketitle

\section{Introduction}

Currently, baryon asymmetry of the Universe and Dark Matter (DM), 
together with nonzero neutrino masses and dark energy, 
represent clear evidences of physics beyond the Standard Model (SM). 
Regarding baryon and DM, they are uncannily similar in two aspects: both are matter 
which experiences gravitational interaction and their cosmic energy densities are of the same order~\cite{Ade:2015xua}
\begin{equation}
\frac{\rho_{\rm DM}}{\rho_{\rm baryon}} 
= \frac{m_X\,Y_X^0}{m_n\,Y_{B_{\rm SM}}^0}
\simeq 5.4\,,\label{eq:ratio_DM_B}
\end{equation}
where $m_X$ and $m_n \sim 1$~GeV are respectively the DM $X$ and the nucleon mass. We also denote $Y_X$ and $Y_{B_{\rm SM}}$ respectively as the number densities of $X$ and the SM baryons $B_{\rm SM}$ normalized to the entropic density $s$, where the superscript `0' indicates their values today. The observed SM baryon abundance today measured by Planck is~\cite{Ade:2015xua}
\begin{equation}
Y_{B_{\rm SM}}^0 = (8.66 \pm 0.09)\times 10^{-11}.
\label{eq:BAU}
\end{equation}
Since the entropy per comoving volume is assumed to be conserved, this value is equal to the SM baryon asymmetry at early time $Y_{B_{\rm SM}}^0 = Y_{\Delta B_{\rm SM}}$ when antibaryons were still in abundance, i.e. at high temperature $T \gg m_n$.
The fact that the SM baryon is maximally asymmetric today is due to the fast nucleon-antinucleon annihilation processes, resulting in all antinucleons to be annihilated, leaving only the access of nucleons due to a baryon asymmetry.
Despite that the DM is `dark', i.e. electric and color charge neutral,  we can suspect that its similarity with the SM baryons does not end here. For instance, DM, like its baryon counterpart, can also be asymmetric. In particular, they can also have some fast interactions to annihilate the symmetric component such that its density today is determined solely by the asymmetric part, i.e. $Y_X^0 = Y_{\Delta X}$.

The idea of an asymmetric DM is a few decades old~\cite{Nussinov:1985xr,Roulet:1988wx,Barr:1990ca} and has seen 
a renewed interest in recent years resulting in a plethora of the new ideas (see some recent review articles~\cite{Davoudiasl:2012uw,Petraki:2013wwa,Zurek:2013wia, Boucenna:2013wba}).
Following ref.~\cite{Bernal:2016gfn}, we will consider a scenario where DM $X$ is a SM singlet but carries either nonzero baryon $B$ and/or lepton number $L$ and couples to the SM particles through the following effective operators
\begin{equation}
\frac{1}{\Lambda^{n-p}} \bar X^2 {{\cal O}}^{\left(n\right)}\,,\label{eq:operator}
\end{equation}
where $\Lambda$ is some effective scale where the operators arise, 
${\cal O}^{\left(n\right)}$ represents SM gauge invariant operator of
mass dimension $n$ with nonzero $B$ and/or $L$ made up of {\it only} the SM fields and $p=1,\,2$ for $X$ being a massive Dirac fermion or a massive complex scalar, respectively.\footnote{A massive Weyl fermion would have to be Majorana and cannot carry an asymmetry, same as a real scalar.}
We consider the case with $X^2$ in eq.~\eqref{eq:operator} to ensure the stability of the DM. This can be realized due to specific $B$ and/or $L$ charge carried by $X$ and a further restriction imposed to forbid nucleon decay: $2\,m_X > m_n$.
Before Electroweak Sphaleron (EWSp) processes freeze out at $T \sim 100$~GeV, the conserved global charge of the SM 
is $B-L$ while after EWSp processes freeze out, the conserved global charges are $B$ and $L$. 
In our scenario, we simply assume a net nonzero charge asymmetry is generated at some high scale and will not discuss its dynamical generation.
By assumption, the \emph{sharing} operator~\eqref{eq:operator} does not violate $B$ nor $L$, and its role is simply to \emph{share} the asymmetry among the SM and the DM sectors. 

In this work, we investigate the Large Hadron Collider (LHC) phenomenology of the sharing scenario of ref.~\cite{Bernal:2016gfn} in order to bound the viable parameter space of the model, taking also into account constraints from Large Electron-Positron collider (LEP).
After a brief review of the asymmetry sharing scenario
in Section~\ref{sec:review}, in Section~\ref{sec:EFT} we  study the constraints on the Effective Field Theory (EFT) operators at the LHC with 13~TeV center of mass energy. 
In Section~\ref{sec:UV_models}, we introduce two representative ultraviolet (UV) complete models. 
In Section~\ref{sec:uvconstr}, we derive collider constraints for these models.
In Section~\ref{sec:new_operator}, we consider a new sharing scenario where the mass of one of the mediators is lighter than 100~GeV.
We conclude in Section~\ref{sec:conclude}.

\section{Review: Baryonic and Leptonic Dark Matter Effective Field Theory} 
\label{sec:review}

For the case where the sharing happens before EWSp processes freeze out, the lowest dimension SM operator 
is of dimension five which carries $B-L = -2$~\cite{Weinberg:1979sa,Weinberg:1980bf}
\begin{equation}
{\cal O}^{\left(5\right)}
 = \epsilon_{ij} \epsilon_{kl}
 ( \bar\ell_{Li}^c \ell_{Lk} ) H_j H_l\,,
  \label{eq:before_EW_op}
\end{equation}
where $\ell_L$ and $H$ are respectively the lepton and the Higgs doublets, and $\psi^c = C \overline \psi^T$ with $C$ the charge-conjugation matrix. We also denote $i$, $j$, $k$, $l$ as the $SU(2)_L$ indices and $\epsilon_{ij}$ the total 
antisymmetric tensor, with $\epsilon_{12}=1$, while the family indices have been suppressed.
The operator ${\cal O}^{\left(5\right)}$
carries $B=0$ and $L=2$, which in turn fixes the baryon and lepton numbers of $X$ according to operator~\eqref{eq:operator} to be  $B_X=0$ and $L_X=1$, respectively.

For the case where the sharing happens after EWSp processes freeze out, there are four dimension six operators 
which carry $B = L = 1$ but $B - L = 0$~\cite{Weinberg:1979sa,Wilczek:1979hc,Abbott:1980zj,Alonso:2014zka}
\begin{eqnarray}
{\cal O}^{\left(6\right){\rm I}}
& = & \epsilon_{ij} ( \bar q_{Li}^c \ell_{Lj} ) 
( \bar d_{R}^c u_{R} ), \label{eq:after_EW_op1}\\
{\cal O}^{\left(6\right){\rm II}}
& = & \epsilon_{ij} ( \bar q_{Li}^c q_{Lj} ) 
(\bar u_{R}^c e_{R} ), \label{eq:after_EW_op2}\\
{\cal O}^{\left(6\right){\rm III}}
& = & \epsilon_{il}\epsilon_{jk} (\bar q_{Li}^c q_{Lj} ) 
(\bar q_{Lk}^{c} \ell_{Ll} ), \label{eq:after_EW_op3}\\
{\cal O}^{\left(6\right){\rm IV}}
& = & (\bar d_{R}^c u_{R} )
(\bar u_{R}^c e_{R} ), \label{eq:after_EW_op4}
\end{eqnarray}
where $q_L$ is a quark doublet, $u_R$ and $d_R$ are respectively 
the up- and down-type quark singlet (family indices are again suppressed),
and the color contractions are implicit. 
All the operators above have both $B$ and $L$ equal to one, which fixes $B_X=L_X=1/2$. 

\begin{figure}[t!]
\centering
\includegraphics[height=5.1cm]{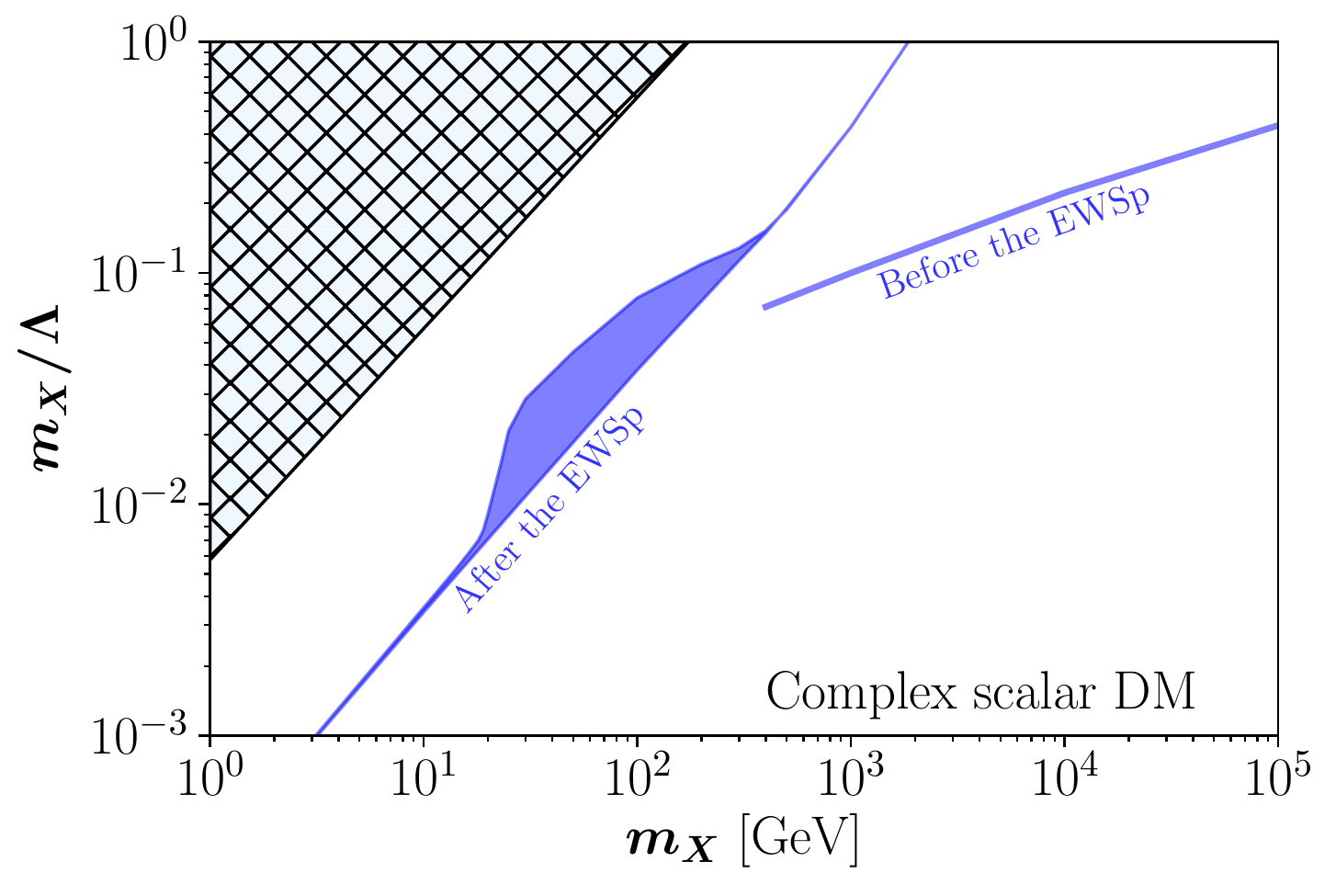}
\includegraphics[height=5.1cm]{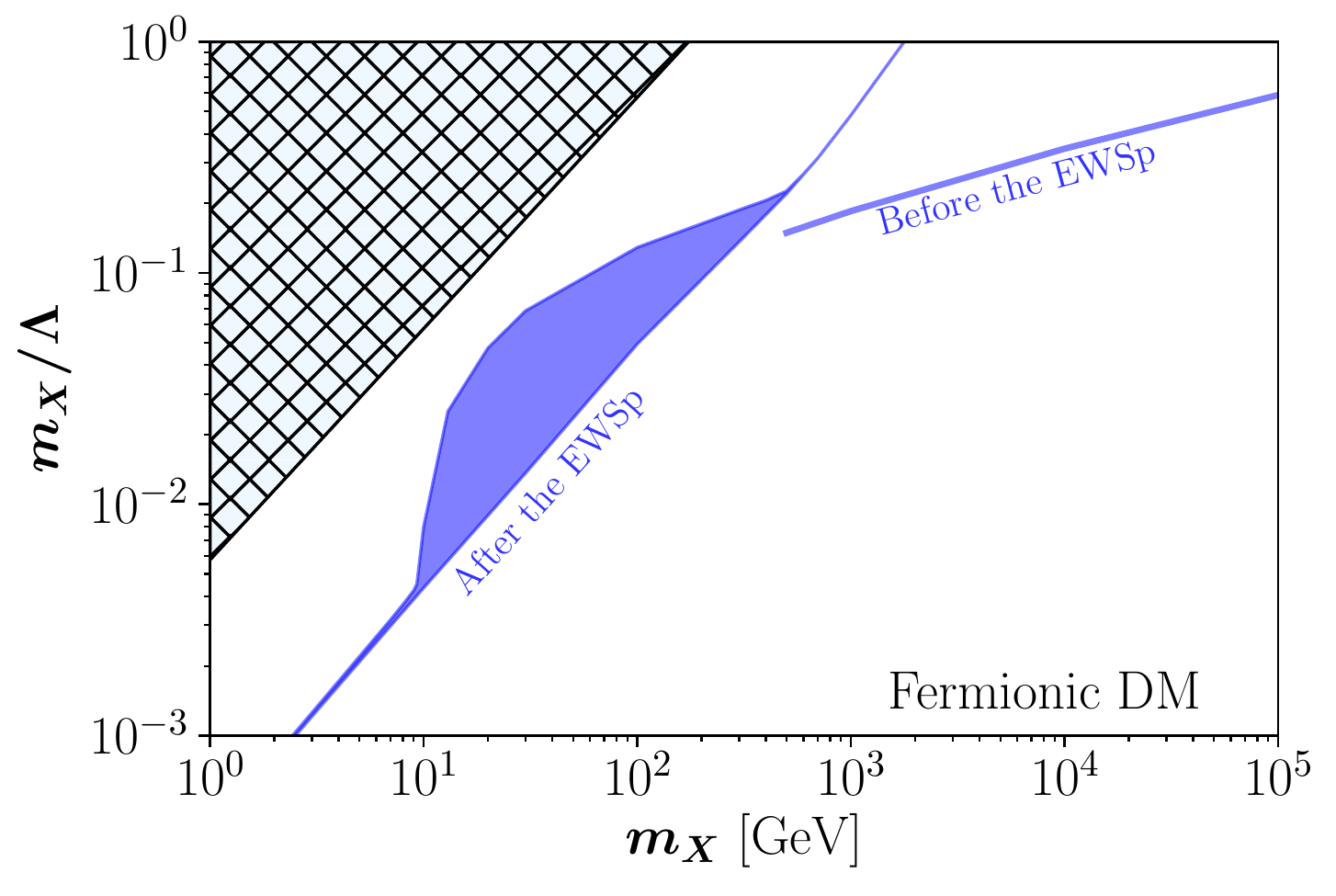}
\caption{Regions of $(m_X,\,\Lambda)$ plane characterized by the fact that the measured baryon asymmetry of the Universe and the DM relic abundance can be reproduced simultaneously, in the case of complex scalar (left panel) and fermionic (right panel) DM, and for the scenarios where the transfer of the asymmetry is efficient before (with operator~\eqref{eq:op5}) and after the freeze-out of the EWSp processes (with operator~\eqref{eq:op6}). Figure adapted from ref.~\cite{Bernal:2016gfn}.
}
\label{mainresults}
\end{figure}

In ref.~\cite{Bernal:2016gfn}, the analysis was carried out considering two realizations of the sharing operator of eq.~\eqref{eq:operator}, where coupling only to the first family SM fermions was assumed. 
For the sharing which took place before the EWSp processes freeze out, the analysis was carried out with eq.~\eqref{eq:before_EW_op} 
while for the sharing which took place after the EWSp processes freeze out, the analysis was done with eq.~\eqref{eq:after_EW_op1}.\footnote{The results from the other three operators~\eqref{eq:after_EW_op2}--\eqref{eq:after_EW_op4} can be obtained through rescaling due to different 
gauge multiplicities as discussed in Appendix~B of ref.~\cite{Bernal:2016gfn}.} In this work we consider the same two realizations of the sharing operators
\begin{eqnarray}
&&\frac{1}{\Lambda^{5-p}} \bar X \bar X \epsilon_{ij} \epsilon_{kl}
 ( \bar\ell_{Li}^c \ell_{Lk} ) H_j H_l\,,
\label{eq:op5}\\
&&\frac{1}{\Lambda^{6-p}} \bar X \bar X 
\epsilon_{ij} (\bar q_{Li}^c \ell_{Lj} ) 
(\bar d_{R}^c u_{R} )
\,,\label{eq:op6}
\end{eqnarray}  
where all the SM fermions refer to the first generation ones.
We assume that the DM abundance today is maximally asymmetric such that given a DM mass, from the observed values eqs.~\eqref{eq:ratio_DM_B} and \eqref{eq:BAU}, one can determine the DM asymmetry $Y_{\Delta X} = Y_{X}^0$. Furthermore, the initial distribution of a fixed total asymmetry is assumed to fully reside either in the SM or the DM sector.
With the above assumptions, the new physics scale $\Lambda$ that appears in eqs.~\eqref{eq:op5} and \eqref{eq:op6} can be determined as a function of the DM mass, namely $\Lambda=\Lambda(m_X)$ by the requirement that the total asymmetry is properly distributed between the SM and the DM sectors, in accordance with the observations. 
The solutions represent the main results of ref.~\cite{Bernal:2016gfn} and are summarized in Fig.~\ref{mainresults}, for both scalar and fermionic DM. The solutions for ``Before the EWSp'' correspond to the operator~\eqref{eq:op5} while those for ``After the EWSp'' correspond to the operator~\eqref{eq:op6}.
For further details, please refer to ref.~\cite{Bernal:2016gfn}.

\section{Constraints on Effective Operators} 
\label{sec:EFT}
In principle, the effective operators in eq.~\eqref{eq:op5} and eq.~\eqref{eq:op6} can be tested at colliders. 
However, one should take care of EFT validity issues because the energies probed at colliders can be larger that the typical mass scale of UV theories that generate those effective operators at low energies. 
In order to deal with this problem we assume the existence of a UV completion that admits a perturbative expansion in its couplings, therefore the effective operator coefficients have the following scaling~\cite{Contino:2016jqw}
\begin{equation}
\frac{1}{\Lambda^{n-p}} = \frac{g^{c-2}}{M^{n-p}}\,,
\end{equation}
where $g$ is the UV dimensionless coupling, $M$ is the mass of the UV states and $c$ the total number of fields entering in the effective operators. 
Perturbativity imposes $g \leq 4\pi$ while the EFT description 
breaks down when the probing energy scale is higher than the mass scale $E > M$. To make sure that the EFT remains valid, in our analysis with simulated events we will fix $g$ and consider only events which satisfy the following condition~\cite{Berlin:2014cfa,Racco:2015dxa}
\begin{equation}
E_{\rm cm} < M = g^{\frac{c-2}{n-p}}\,\Lambda\, ,
\label{eq:eftcut}
\end{equation}
where $E_{\rm cm}$ is the partonic center of mass energy.

\subsection{Energetic Jet(s) + MET Searches at LHC}
\begin{figure}[t!]
\centering
\includegraphics[height=4cm]{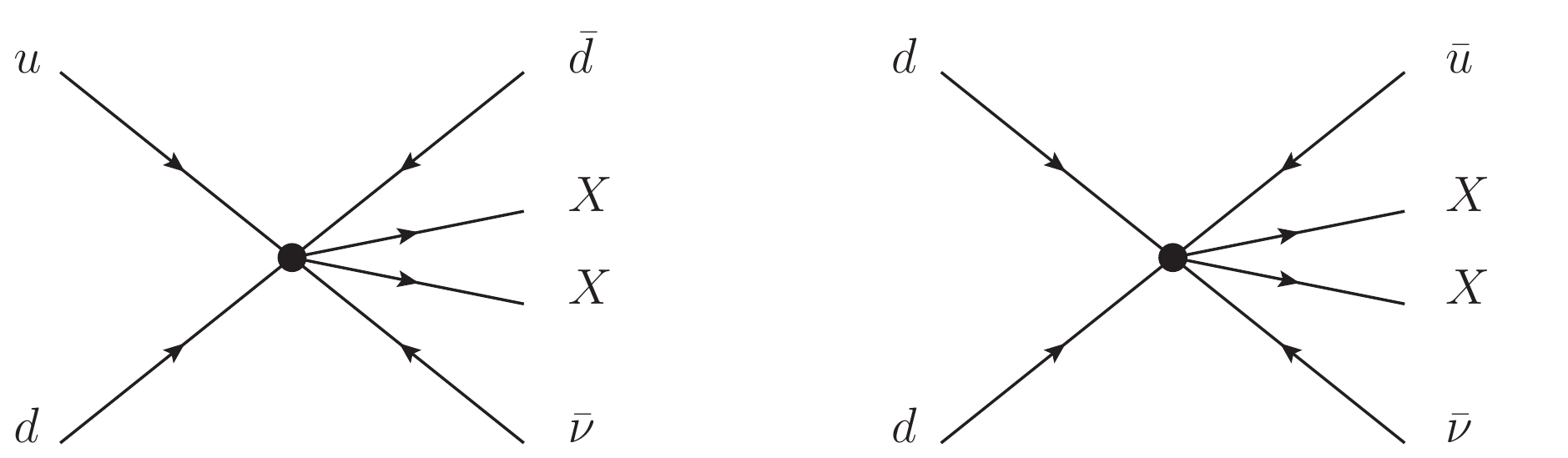}
\caption{Feynman diagrams for $pp\to j\nu XX$ production at the LHC from operator \eqref{eq:op6}.}
\label{diagram1}
\end{figure}

LHC searches for new phenomena in events featuring energetic jet(s) + large MET~\cite{Aaboud:2016tnv,Aaboud:2017buf}  can be used to bound the operator in eq.~\eqref{eq:op6} because this operator gives rise to the process
\begin{equation}
pp\to j \bar\nu XX
\label{eq:monojet}
\end{equation}
which contributes to jet(s) + MET final state as shown in Fig.~\ref{diagram1}. 
We have included also the conjugate process $p p \to j \nu \bar X \bar X$ which is much more suppressed due to proton Parton Distribution Function (PDF), dominated by quarks instead of antiquarks.
In order to perform this sensitivity study we implement the effective operator~\eqref{eq:op6} in \texttt{FeynRules}~\cite{Alloul:2013bka} for both scenarios of scalar and fermionic DM and create the \texttt{UFO}~\cite{Degrande:2011ua} interface which is then used in \texttt{MadGraph5}~\cite{Alwall:2011uj} to generate $pp\to j \bar\nu XX$ events at $\sqrt{s}=13$~TeV. We run the event simulations for different values of $\Lambda$ and $m_X$ for the viable parameter space
of the case where the transfer of the asymmetry is efficient after the EWSp processes freeze-out (see Fig.~\ref{mainresults}). EFT validity issues affecting the generated events has been addressed as discussed at the beginning of Section~\ref{sec:EFT}, namely
by assuming a value for the UV coupling $g$ ($\leq 4\pi$) and retaining at generator level only events that satisfy the following EFT validity cut (c.f. eq.~\eqref{eq:eftcut})
\begin{equation}\label{eftbound}
\sqrt{s} < \Bigg\{
\begin{array}{cc}
g\,\Lambda\;\;&\;\;\mbox{for scalar DM,}\\
g^{4/5}\Lambda\;\;&\;\;\text{for fermionic DM.}
\end{array}
\end{equation}
Events that pass the EFT validity cut are then showered using \texttt{PYTHIA8}~\cite{Sjostrand:2014zea}.

\begin{figure}[t!]
\centering
\includegraphics[height=5.3cm]{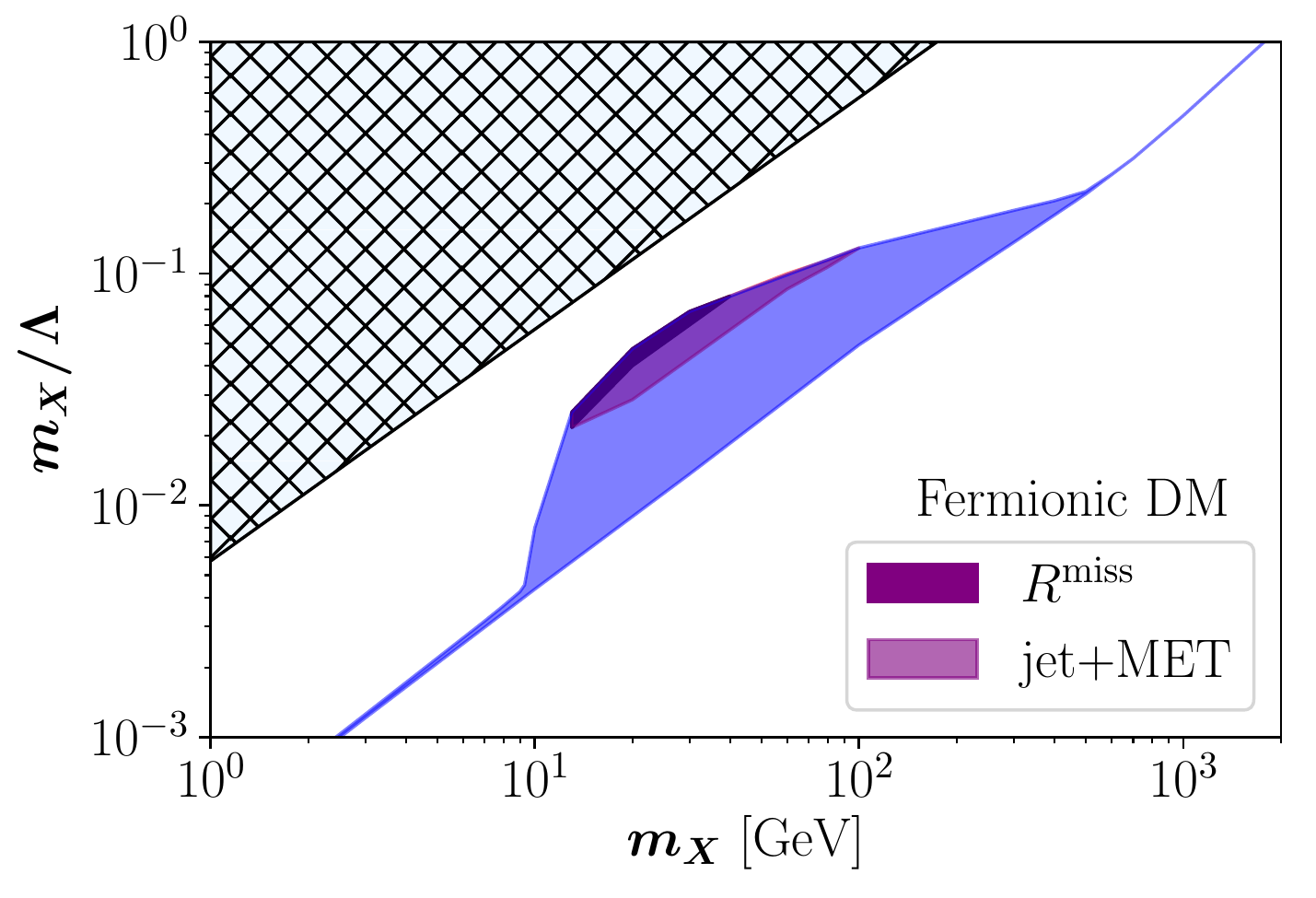}
\caption{Exclusion regions for the fermionic DM case obtained from the LHC recasting of the energetic jet + MET searches~\cite{Aaboud:2016tnv} (pink) and from $R^{\rm miss}$ measurements~\cite{Aaboud:2017buf} (purple), at $\sqrt{s}=13$~TeV and integrated luminosity of 3.2~fb$^{-1}$.}
	\label{monojetbounds}
\end{figure}

The generated events are passed to the \texttt{RIVET}~\cite{Buckley:2010ar} analysis \texttt{ATLAS\_2016\_I1452559} based on ref.~\cite{Aaboud:2016tnv}. Events have been selected using the following criteria (for complete details see ref.~\cite{Aaboud:2016tnv}):
\begin{itemize}
\item $E_{\rm T}^{\rm miss}> 250$~GeV,
\item leading (highest) $p_{\rm T}$ jet with $p_{\rm T} > 250$ GeV and $|\eta|<2.4$ in the final state,
\item a maximum of four jets with $p_{\rm T} > 30$~GeV and $|\eta|<2.8$ are allowed,
\item a separation in the azimuthal plane of $\Delta\phi({\rm jet},\,p_{\rm T}^{\rm miss}) > 0.4$
between the missing transverse momentum direction and each selected jet is required,
\item events with identified muons with $p_{\rm T} > 10$~GeV or electrons with $p_{\rm T} > 20$~GeV in the final state are vetoed.
\end{itemize}
The $E_{\rm T}^{\rm miss}$ is reconstructed using all energy deposits in the calorimeter up to pseudorapidity $|\eta|= 4.9$. The most sensitive signal region turned out to be the IM7 inclusive region, characterized by the additional requirement $E_{\rm T}^{\rm miss}> 700$~GeV. The \texttt{RIVET} analysis provides the number of expected events in the IM7 region surviving the selection criteria at 3.2~fb$^{-1}$ which have been compared with the observed 95\% CL upper limits on the number of signal events taken from Table~9 of ref.~\cite{Aaboud:2016tnv}. A point in the $(m_X,\Lambda)$ plane is considered excluded if the number of expected events is bigger than 61. We find that no excluded regions are present in the scalar DM case, while there are some excluded regions in the fermionic case (assuming $g=4\pi$) for DM mass values ranging from 10 to 100~GeV and those results are shown in Fig.~\ref{monojetbounds}. 

On the same events we run also the \texttt{RIVET} analysis \texttt{ATLAS\_2017\_I1609448} based on ref.~\cite{Aaboud:2017buf} that presents a measurement of differential observables that are sensitive to the anomalous production of events containing one or more hadronic jets with high transverse momentum, produced in association with a large $p_T^{\rm miss}$ at 3.2~fb$^{-1}$. These observables are constructed from a ratio of cross-sections
\begin{equation}
R^{\rm miss}=\frac{\sigma_{\rm fid}(p_T^{\rm miss}+\rm jets)}{\sigma_{\rm fid}(l^+l^-+\rm jets)}
\end{equation}
defined in a fiducial phase space and we have used the measured $R^{\rm miss}$ as a function of $p_T^{\rm miss}$ in the $n\geq 1$ jet region (for more details see ref.~\cite{Aaboud:2017buf}). For each point in the $(m_X,\Lambda)$ plane, the \texttt{RIVET} analysis provides us with the SM plus signal expectations of this $R^{\rm miss}$ observable. We have compared the expectations with experimental data and performed a binned $\chi^2$ analysis taking into account also correlations (provided in \texttt{HEPData}~\cite{Maguire:2017ypu}) in order to determine which points are excluded or not. We find that no excluded regions are present in the scalar DM case, while there are some excluded regions in the fermionic case (assuming $g=4\pi$) for DM mass values ranging from 10 to 50~GeV. These results are shown in Fig.~\ref{monojetbounds}.

Looking at the results of Fig.~\ref{monojetbounds}, it turns out that the $R^{\rm miss}$ analysis in the $n\geq 1$ jet region of ref.~\cite{Aaboud:2017buf} is less sensitive than the IM7 inclusive region analysis of ref.~\cite{Aaboud:2016tnv}. The results have been presented only for the case $g=4\pi$ which is expected to be the scenario that provides the largest exclusion limits. For the IM7 analysis we have checked that by reducing the value of $g$, the exclusion region shrinks and below $g=7$,
we do not have sensitivity to exclude any part of the viable parameter space.
This is due to the fact that smaller $g$ values (corresponding to smaller mediator mass scales) force a larger number of events to be rejected in order to fulfill the EFT validity cut of eq.~\eqref{eq:eftcut}, thus reducing the sensitivity. 
In other words, EFT only allows us to constraint scenario with rather large coupling $g \gtrsim 7$.

\subsection{Other LHC Signatures}
LHC measurements of same sign di-boson production in association with two jets $p p  \to W^\pm W^\pm j j\to l^\pm \nu l^\pm \nu j j$  can be used to set constraints on the operator in eq.~\eqref{eq:op5} given the fact that it gives rise to the following processes 
\begin{equation}
p p \to j j \ell^{\pm} \ell^{\pm} \overset{(-)}{X} \overset{(-)}{X},
\label{eq:other1}
\end{equation}
which contributes to the same final state consisting of two jets, two same sign leptons and MET. 
Notice that the process $p p \to j j \ell^{+} \ell^{+} X X$ dominates over $p p \to j j \ell^{-} \ell^{-} \bar X \bar X$ due to proton PDF, dominated by quarks instead of antiquarks.
However, for all the $(m_X,\,\Lambda)$ points relevant for the sharing mechanism, the cross section turns out to be several orders of magnitude smaller than the SM background and therefore we have no sensitivity in this channel.

The operator~\eqref{eq:op6} can also be constrained by the LHC measurements of $W$ production cross sections in association with jets, in particular by the $W$+ 1 jet production $p p \to W^\pm j \to e^\pm \nu j$. It also generates the following process
\begin{equation}
p p \to j e^+ X X,
\label{eq:other2}
\end{equation}
which contributes to the same final state consisting of one positron, one jet and MET similar to the diagram of Fig.~\ref{diagram1} but with $\nu$ replaced by $e$. Similar to process \eqref{eq:other1}, due to proton PDF, we have that $p p \to j e^+ X X$ dominates over $p p \to j e^- \bar{X} \bar{X}$. These two processes can in principle be distinguished and the asymmetry represents a distinctive signature of the sharing scenario.
We have generated $pp\to j e^+ XX$ events at $\sqrt{s}=7$~TeV and run the \texttt{RIVET} analysis \texttt{ATLAS\_2014\_I1319490\_EL}, based on ref.~\cite{Aad:2014qxa}. We found no sensitivity in this channel. However we expect to get sensitivity from future measurements of $W$+1 jet in the electron channel at 13~TeV.

\section{UV Models} 
\label{sec:UV_models}
After investigating the constraints on the effective operators taking into account the validity of EFT, we will now consider representative UV models giving rise at low energies to the effective operators in eqs.~\eqref{eq:op5} and~\eqref{eq:op6}. 

\subsection[A UV Model for $\bar X \bar X{ {\cal O}}^{(5)}$]{A UV Model for $\boldsymbol{\bar X \bar X{ {\cal O}}^{(5)}}$} 
\label{sec:UV_O5}

Here we consider a simple UV model that can generate the operator in eq.~\eqref{eq:op5} at low energies. In addition to a DM field $X$ with mass $m_X$, we introduce a heavy complex scalar $\phi$ with mass $m_\phi$ and a heavy Dirac fermion $\psi=\psi_L+\psi_R$ with mass $m_\psi$ which are both singlets under the SM gauge symmetry but are charged under the lepton number $U(1)_L$ as follows: 
\begin{equation}
\frac{L(\phi)}{2}=L(\psi)=L(X)=1.
\end{equation}
The Lagrangian of the model contains the following terms\footnote{For simplicity, we consider couplings only to first generation SM leptons.}
\begin{equation}\label{lagbl}
-{\cal L}\supset m_\psi\,\bar\psi\,\psi+m_\phi^2\,\phi^*\,\phi+ \left(\lambda\,\overline{\psi_R}\,\ell_L\epsilon\,H+\eta\,\overline{(\psi^{c})_R}\,\psi_L\,\phi^{*}+\mu\,m_\phi^{p-1}\,\phi^{*}\,X^{2}+{\rm h.c.}\right),
\end{equation}
where $\lambda$, $\eta$ and $\mu$ are dimensionless couplings. 
At energies $E\ll m_{\phi},\,m_{\psi}$, we can perform a tree-level matching and write the Wilson coefficient of the operator in eq.~\eqref{eq:op5} in terms of the parameters of the UV model in eq.~\eqref{lagbl}
\begin{equation}
\frac{1}{\Lambda^{5-p}}  =  \frac{\lambda^2\,\eta\,\mu}{m_{\phi}^{3-p}m_{\psi}^{2}}.
\label{eq:op5_matching}
\end{equation}
From the Lagrangian in eq.~\eqref{lagbl}, after the Higgs doublet acquires a vacuum expectation value $v \equiv \left<H\right>\simeq 174$~GeV, we have that one linear combination of $\nu_L$ and $\psi_L$ couples to $\psi_R$ forming a massive state, while the orthogonal linear combination remains massless. They are respectively given by:\footnote{The massless $\tilde \nu_L$ corresponds to the SM neutrino and its tiny mass can be generated through other mechanism like the seesaw mechanism.}
\begin{eqnarray}
\tilde \psi_L&=&+\cos\theta\,\psi_L+\sin\theta\,\nu_L ,\nonumber\\
\tilde \nu_L&=&-\sin\theta\,\psi_L+\cos\theta\,\nu_L ,
\end{eqnarray}
where
\begin{equation}
\tan\theta = \frac{\lambda\,v}{m_\psi}.\label{eq:mixing}
\end{equation}
The massive state $\tilde \psi$ has a mass $m_{\tilde \psi}=\sqrt{m_\psi^2+\lambda^2\,v^2}$. Assuming small mixing angle $\theta \ll 1$, the matching condition \eqref{eq:op5_matching} can be rewritten as
\begin{equation}
\frac{1}{\Lambda^{5-p}}  =  \frac{\eta\,\mu \sin^2\theta}
{m_{\phi}^{3-p} v^2}.
\label{eq:op5_matching2}
\end{equation}
There exists heavy neutral lepton constraints on $m_\psi$ and $\sin\theta$~\cite{deGouvea:2015euy}. 
For $m_{\tilde \psi} \gtrsim 100$ GeV, lepton universality and invisible $Z$ width bound $\sin\theta \lesssim 10^{-1}$.
From the solutions (``Before the EWSp'') from Fig.~\ref{mainresults}, we have $\Lambda \gtrsim 10$~TeV while the validity of our EFT approach to sharing scenario requires $m_{\tilde \psi}$,\,$m_\phi \gtrsim m_X \sim 400$~GeV~\cite{Bernal:2016gfn}. We can easily choose the couplings which satisfy eq.~\eqref{eq:op5_matching2} as well as the mixing constraints. 

At an electron-positron collider, $\psi$ can be produced through mixing \eqref{eq:mixing} with the SM neutrino, giving rise to mono-photon signature: $e^+ e^- \to {\tilde \psi} \bar {\tilde \psi} \gamma$. In addition to the suppression from mixing, the required center of mass energy $\sqrt{s} > 2\,m_{\tilde \psi}$ makes this scenario unlikely to be probed in the near future.

\subsection[A UV Model for $\bar X \bar X {\cal O}^{(6)}$]{A UV Model for $\boldsymbol{\bar X \bar X {\cal O}^{(6)}}$}
\label{sec:UV_O6}
\begin{figure}[t!]
\centering
\includegraphics[height=4cm]{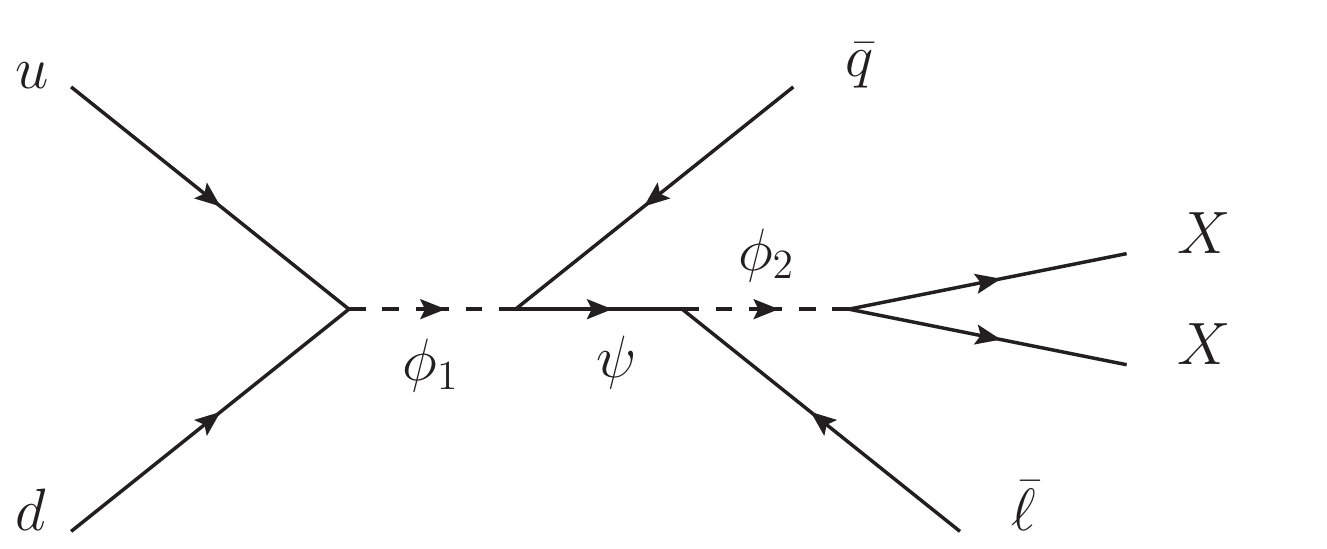}
\caption{Example of tree level diagram contributing to the matching condition of eq.~\eqref{match06} which arises from the UV model of Lagrangian~\eqref{lag6}.
}
\label{diagrammatching}
\end{figure}

Here we will consider a simple UV model which is able to reproduce the operator in eq.~\eqref{eq:op6} at low energies. Therefore, in addition to a DM field $X$ with mass $m_X$,  we introduce two heavy complex scalars $\phi_1$ and $\phi_2$ with masses $m_{\phi_1}$ and $m_{\phi_2}$ and a heavy Dirac fermion $\psi=\psi_L+\psi_R$ with mass $m_\psi$. 
We have in mind that all these mediators have mass larger than $\sim 100$~GeV such that the analysis of ref.~\cite{Bernal:2016gfn} with EFT remains valid. For the sake of completeness, in Section~\ref{sec:new_operator}, we explore the possibility that some of them have a smaller mass.
These new fields are charged under the SM symmetry group $SU(3)_{c}\times SU(2)_{L}\times U(1)_{Y}$, but also under the baryon and lepton number symmetry $U(1)_L$ and $U(1)_B$. The fermion $\psi$ being vector-like under the SM gauge interactions does not introduce gauge anomalies. The quantum numbers of these fields are listed in Table~\ref{tab:XXdim6_model}.
\begin{table}[t!]
\begin{center}
\begin{tabular}{|c||c|c|c|c|c|}
\hline 
{\bf Fields} & $\bf SU(3)_{c}$ & $\bf SU(2)_{L}$ & $\bf U(1)_{Y}$ & $\bf U(1)_{B}$ & $\bf U(1)_{L}$\tabularnewline
\hline 
\hline 
$\boldsymbol\phi_{1}$ & 3 & 1 & $\frac{1}{3}$ & $\frac{2}{3}$ & 0\tabularnewline
\hline 
$\boldsymbol\phi_{2}$ & 1 & 1 & 0 & 1 & 1\tabularnewline
\hline 
$\boldsymbol\psi$ & 1 & 2 & $\frac{1}{2}$ & 1 & 0\tabularnewline
\hline 
$\boldsymbol X$ & 1 & 1 & 0 & $\frac{1}{2}$ & $\frac{1}{2}$\tabularnewline
\hline 
\end{tabular}
\caption{New fields of the UV model with their charges.\label{tab:XXdim6_model}}
\end{center}
\end{table}
The Lagrangian of the model contains the following terms 
\begin{eqnarray}\label{lag6}
-{\cal L}&\supset & m_\psi \bar \psi \psi+m_{\phi_1}^2 \phi_1^*\phi_1 +m_{\phi_2}^2 \phi_2^*\phi_2\nonumber \\
&&+ \left[\lambda\,\overline{d_R}(u^{c})_L\phi_{1}+\eta\,\overline{\psi}_R
q_L\phi_{1}+\zeta\,\overline{\ell}_L \epsilon(\psi^{c})_R\phi_{2}
+\mu\,m_{\phi_2}^{p-1}\phi_{2}^{*}X^{2}+{\rm h.c.}\right],
\end{eqnarray}
where $\lambda$, $\eta$, $\zeta$ and $\mu$ are dimensionless couplings. The color contractions in eq.~\eqref{lag6} are left implicit. At low energy $E\ll m_{\phi_{1}}$, $m_{\phi_{2}}$, $m_{\psi}$, we can perform a tree level matching and write the Wilson coefficient of the operator in eq.~\eqref{eq:op6} in terms of the parameters of the UV model
\begin{equation}\label{match06}
\frac{1}{\Lambda^{6-p}}  =  \frac{\lambda\,\eta\,\zeta\,\mu}{m_{\phi_{1}}^{2}m_{\phi_{2}}^{3-p}m_{\psi}}.
\end{equation}
An example of tree level matching diagram is shown in Fig.~\ref{diagrammatching}, in the limit $E\ll m_{\phi_{1}}$, $m_{\phi_{2}}$, $m_{\psi}$ this reduces to the diagram in Fig.~\ref{diagram1}.

\section{Constraints on UV Models}
\label{sec:uvconstr}

Here we will derive constraints on the UV models proposed in the previous section. In particular, we will demonstrate that the bounds derived can change based on assumptions on the couplings and mass of the mediators of these models.

\subsection{Di-jet Searches at LHC}
\label{sec:di-jet}
Di-jets searches at LHC~\cite{Khachatryan:2015dcf} can be used to bound some regions of the parameter space of the UV model described in Section~\ref{sec:UV_O6}. Indeed, due to the first term of eq.~\eqref{lag6}, we can have a production of a di-quark resonance $pp\to\phi_{1}$ which can decay back into di-jets $\phi_1\to u d$. This process depends on $\lambda$ and $m_{\phi_{1}}$. In order to set a bound of $m_{\phi_1}$ as function of the coupling $\lambda$ we have implemented the model in \texttt{FeynRules} and created the \texttt{UFO} interface which has been used in \texttt{MadGraph5} to generate $pp\to \phi_{1}\to j j$ events at $\sqrt{s}=13$~TeV.

\begin{figure}[t!]
\centering
\includegraphics[height=5.3cm]{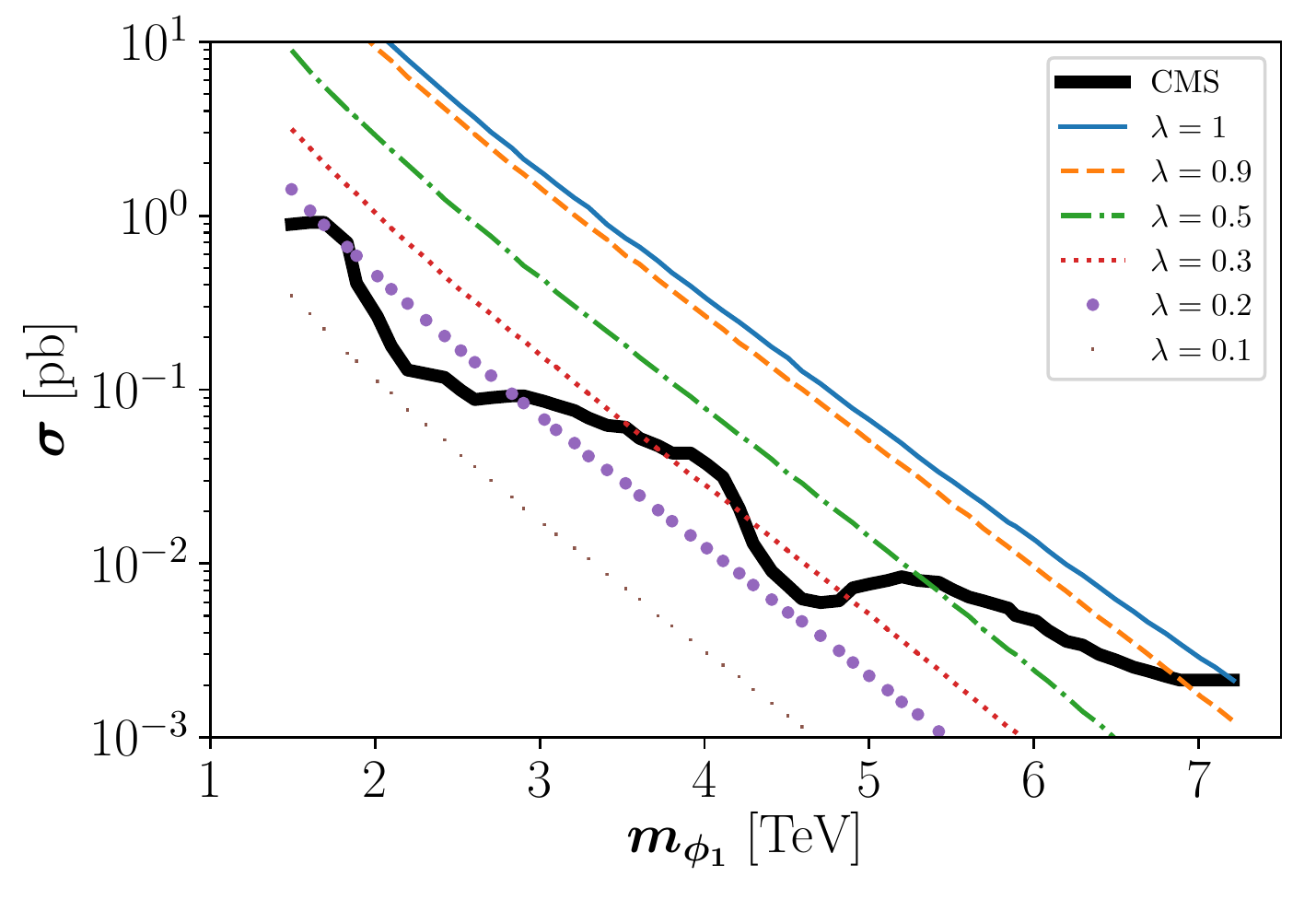}
\caption{The solid black curve represents the CMS~\cite{Khachatryan:2015dcf} observed 95\%~CL upper limits on production cross section times  branching fraction as function of di-jet resonance mass.
Those limits are compared to the predicted cross sections of $pp\to \phi_1\to jj$ computed in our UV model 
 as function of  $m_{\phi_1}$ and for different values of the coupling $0.1<\lambda<1$. These predictions are represented by the colored curves.}
\label{dijets}
\end{figure}

We computed the partonic cross section for $\phi_1$ production and decay $pp\to \phi_1\to jj$ as function of  $m_{\phi_1}$ in the range $1$~TeV $<m_{\phi_1}<7$~TeV, for different values of the coupling $0.1<\lambda<1$. We then compared our cross section estimations with the 95\%~CL experimental curve obtained by CMS in ref.~\cite{Khachatryan:2015dcf}. This result is shown in Fig.~\ref{dijets}
and from the plot we can read off the (strongest) bound on the
mass $m_{\phi_1}$, which is for $\lambda = 0.5$:
\begin{equation}\label{bounddijet}
m_{\phi_1}\gtrsim 5.4~{\rm TeV}.
\end{equation}

From the matching condition eq.~\eqref{match06},
the \emph{simplest} version of the UV model consists in assuming universal couplings
\begin{equation}\label{eq:simplest1}
\lambda=\eta=\zeta=\mu \equiv g,
\end{equation}
and universal masses
\begin{equation}
m_{\phi_{1}}=m_{\phi_{2}}=m_{\psi} \equiv M.
\end{equation}
In this case we have
\begin{equation}\label{matchsimp}
\frac{1}{\Lambda^{6-p}}  =  \frac{g^{4}}{M^{6-p}}\,.
\end{equation}
If we interpret the di-jet result in terms of the \emph{simplest} version of the model, 
using eqs.~\eqref{bounddijet} and~\eqref{matchsimp}, with $g = 0.5$, we have
\begin{equation}\label{boudsimp}
\Lambda \gtrsim\Bigg\{
\begin{array}{cc}
11\;{\rm TeV}\;\;&\;\;\text{for scalar DM,}\\
9.5\;{\rm TeV}\;\;&\;\;\mbox{for fermionic DM.}
\end{array}
\end{equation}
\begin{figure}[t!]
\centering
\includegraphics[height=5.2cm]{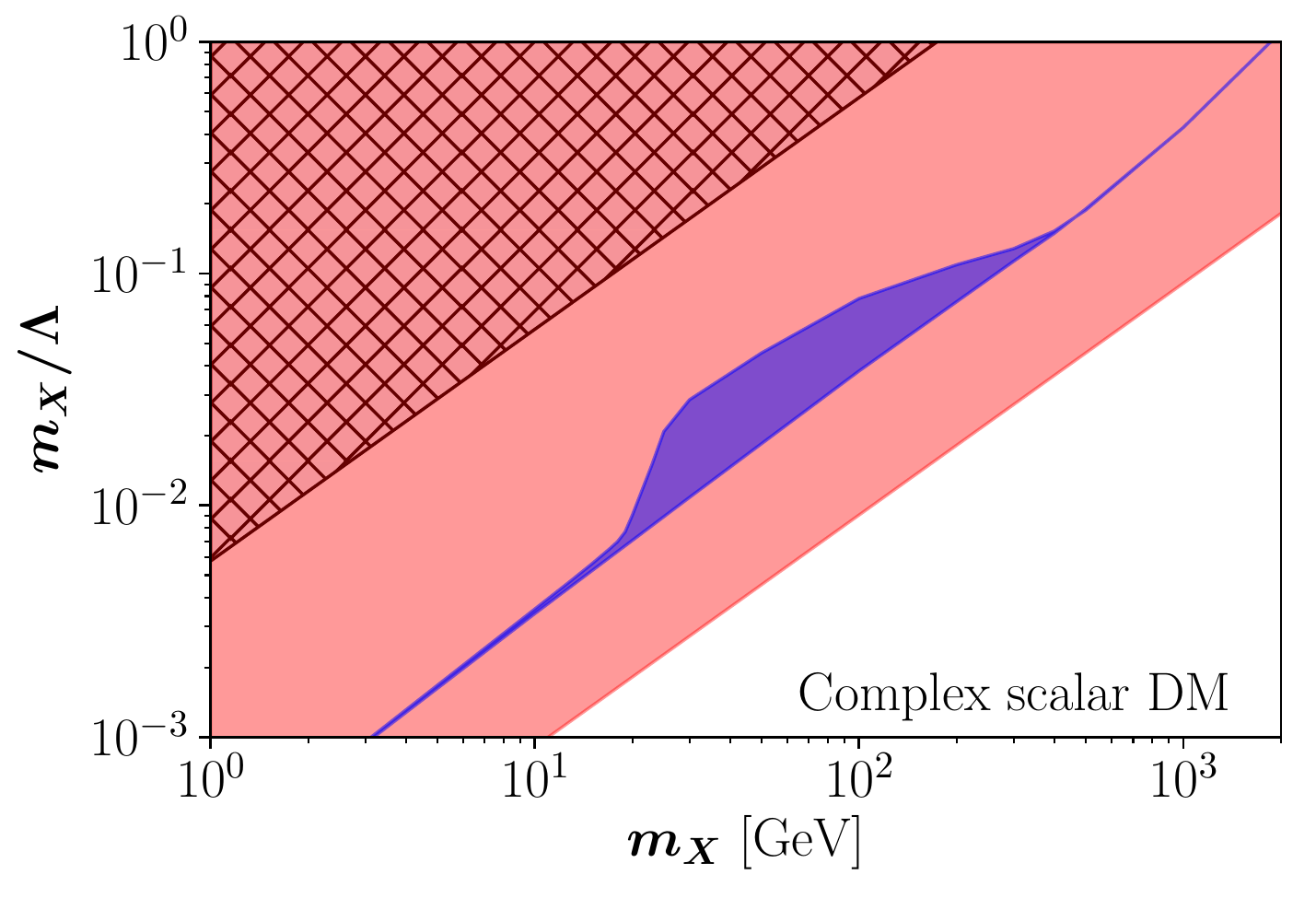}
\includegraphics[height=5.2cm]{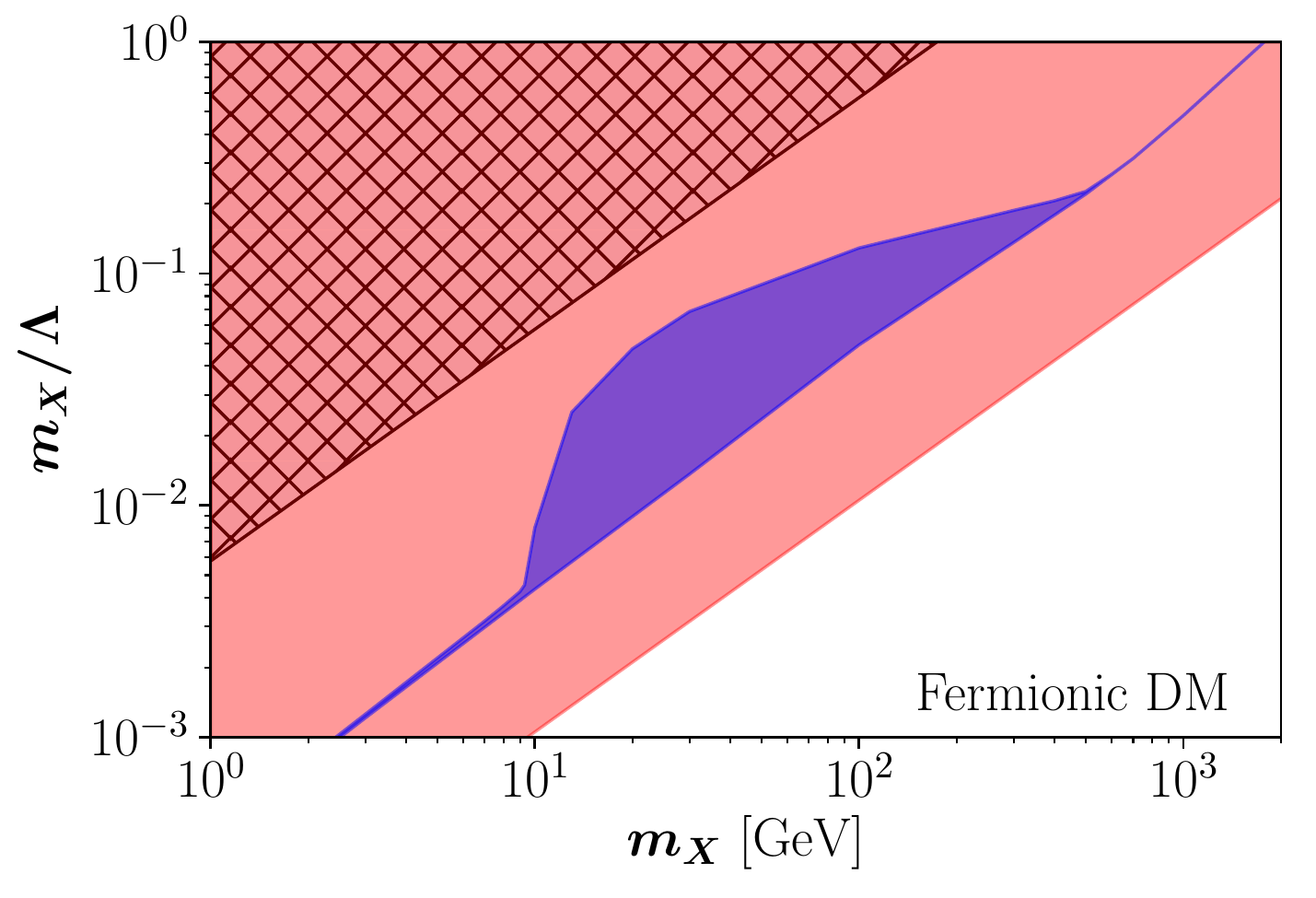}
\caption{The red shaded area represents the excluded region in the $(m_X,\,m_X/\Lambda)$ plane for scalar DM (left) and fermionic DM (right), in the simplest realization of the UV model (eqs.~\eqref{eq:simplest1}-\eqref{matchsimp}).}
\label{okmola2}
\end{figure}
Notice that this result is independent of the mass of the DM. Eq.~\eqref{boudsimp} completely rules out this particular realization of the sharing scenario; from Fig.~\ref{okmola2} we can see that there is no point in the $(m_X/\Lambda,\,m_X)$ plane that satisfies both the sharing requirements and  eq.~\eqref{boudsimp}. 

Since the bounds on the mass $m_{\phi_1}$ of new color particle $\phi_1$ and coupling $\lambda$ are quite stringent, in the following we consider the \emph{next-to-simplest} version of the UV model which is characterized by the assumptions  
\begin{equation}
\eta=\zeta=\mu \equiv g\neq \lambda
\end{equation}
and
\begin{equation}
m_{\phi_{2}}= m_{\psi}\equiv M\neq m_{\phi_{1}}\,.
\end{equation}
In this case, the matching condition becomes
\begin{equation}\label{matchnext}
\frac{1}{\Lambda^{6-p}}  =  \frac{\lambda\,g^3}{m_{\phi_{1}}^2\,M^{4-p}}\,.
\end{equation}
In terms of this \emph{next-to-simplest} model
we have
\begin{equation}\label{boundnext}
\Lambda \gtrsim\Bigg\{
\begin{array}{cc}0.9
\left(\frac{1}{g}\right)^{\frac{3}{4}}
\left(\frac{M}{110~{\rm GeV}}\right)^{\frac{1}{2}}
\;{\rm TeV}\;\;&\;\;\text{for scalar DM,}\\
0.6
\left(\frac{1}{g}\right)^{\frac{3}{5}}
\left(\frac{M}{110~{\rm GeV}}\right)^{\frac{3}{5}}
\;{\rm TeV}\;\;&\;\;\mbox{for fermionic DM.}
\end{array}
\end{equation}
The di-jet production cross section by itself does not allow to extract a bound on $\Lambda$ as function of $m_X$ for this next-to-simplest model. In order to do this we need to consider another observable that allows to constrain $M$ and $g$ independently. This observable will be the photon production in association with MET at LEP and will be discussed in the next section.

\subsection{Photon Production in Association with MET at LEP}
\label{sec:LEP2}
\begin{figure}[t!]
\centering
\includegraphics[height=4cm]{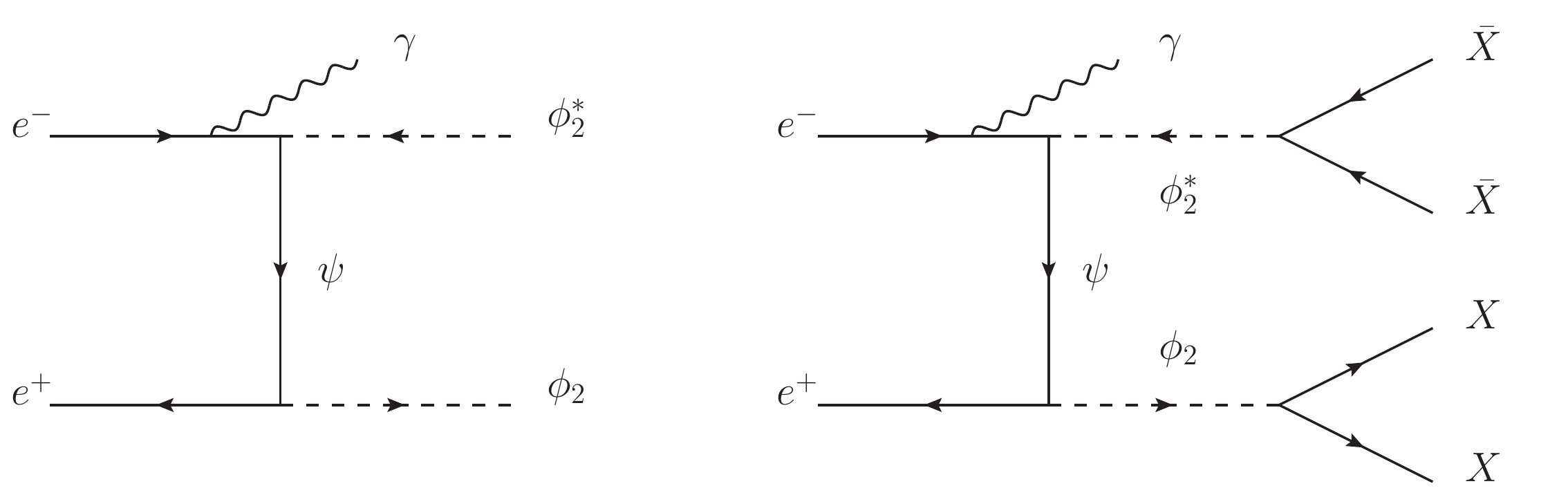}
\caption{Feynman diagrams for $e^+ e^- \to \phi_2 \phi_2^* \gamma$ and $e^+ e^- \to XX\bar X \bar X \gamma$ at LEP.}
\label{diagram2}
\end{figure}

Photon production in association with MET at LEP~\cite{Abdallah:2003np} can be used to bound a region of the parameter space of the UV model of Section~\ref{sec:UV_O6}, which is complementary to the region bounded by di-jet searches at LHC. Here we consider \emph{next-to-simplest} model, eq.~\eqref{matchnext}, and we have fixed $M = 110$~GeV to be consistent with the requirement $m_\psi=M>100$~GeV from not having produced a pair of electroweak-charged $\psi$ particles at LEP~\cite{Achard:2001qw}.
In principle there could also be bounds from multi-lepton (three or more) searches~\cite{Chatrchyan:2014aea,Aad:2014hja} and $W^+ W^-$ production at LHC~\cite{Aad:2016wpd} but they do not apply to pair production of $\psi$, which gives rise to two-lepton final states. Therefore we have the following process contributing to mono-photon + MET production: $e^+ e^- \to XX\bar X \bar X \gamma$.
This process depends on $M$, $m_X$ and $g$; the representative Feynman diagram is shown in  Fig.~\ref{diagram2}.
The observable used to constrain the model is the photon energy spectrum measured by the DELPHI collaboration~\cite{Abdallah:2003np, Abdallah:2008aa}. 

\begin{figure}[t!]
\centering
\includegraphics[height=5.3cm]{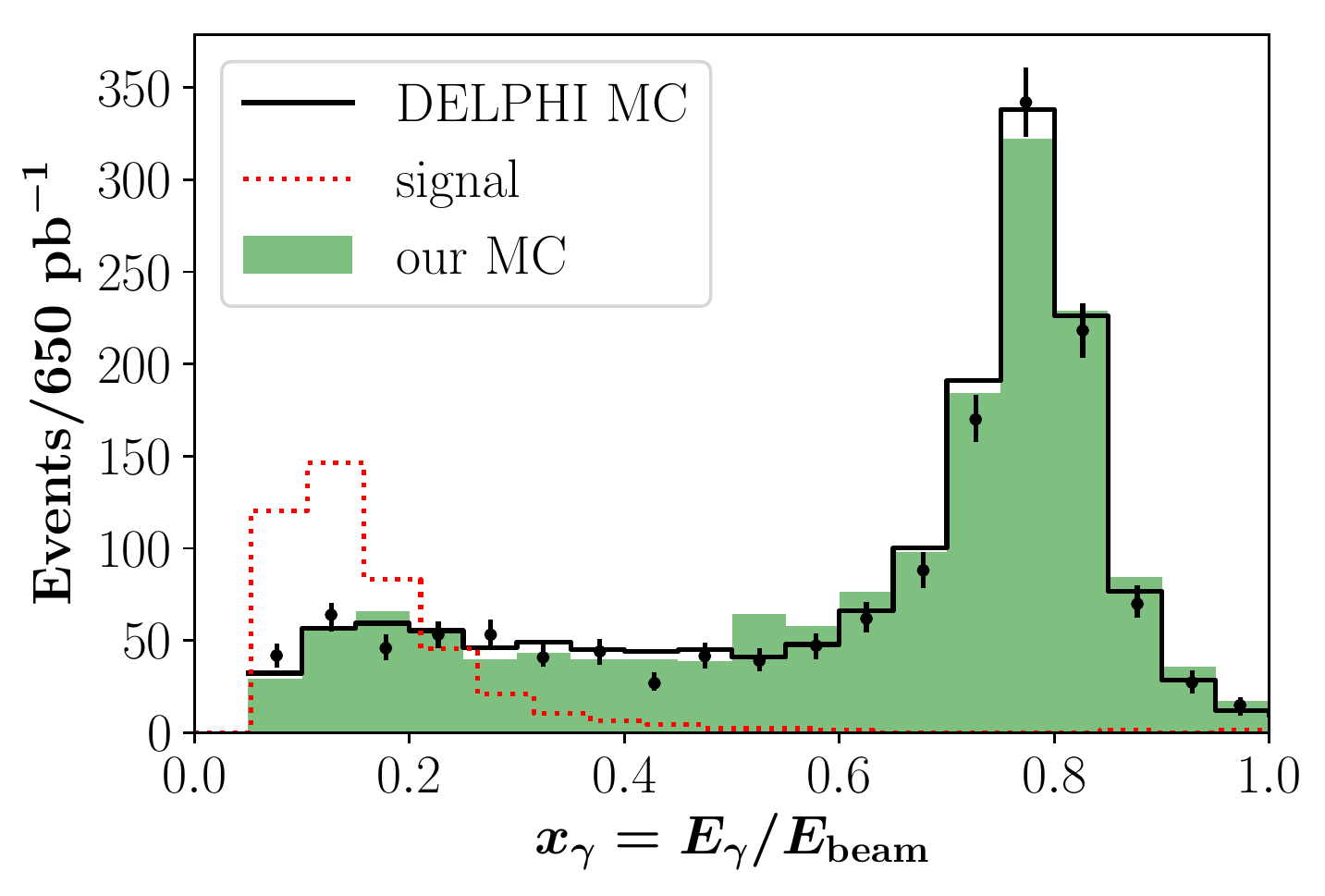}
\caption{Distribution of normalized photon energy in single-photon events at DELPHI.
The data (black dots with error bars) as well as the DELPHI Monte Carlo (black histogram) and our \texttt{Whizard} simulation (green histogram) are shown.
The peak at $x_\gamma\sim 0.8$ corresponds to the process $e^+e^-\to\gamma Z\to\gamma\nu\bar\nu$ with an on-shell $Z$. 
The red dotted histogram corresponds to the signal photon spectrum for the process $e^+e^-\to XX\bar X\bar X\gamma$, for $g=3$ and $M=20$~GeV.}
\label{Delphi}
\end{figure}

We first simulate the SM background at particle level using \texttt{Whizard}~\cite{Kilian:2007gr}, which allows us to properly take into account initial state radiation. We then simulate  the detector response by following the procedure implemented in ref.~\cite{Fox:2011fx} and we check that the simulated background distribution is in agreement with data, as shown in Fig.~\ref{Delphi}. The data was taken at center of mass energies between 180~GeV and 209~GeV, but since the measurement takes into account the relative photon energy $E_\gamma/E_{\rm beam}$, we can make the simplifying assumption that all data was taken at an energy of 100~GeV per beam.

We simulate the signal events 
$e^+ e^- \to XX\bar X \bar X \gamma $ at particle level for different values of $M$, $m_X$ and $g$ using \texttt{Whizard} and we construct the photon energy spectrum by taking into account the detector response. In Fig.~\ref{Delphi} the 
distribution of normalized photon energy in single-photon events at DELPHI is shown.
The data (black dots with error bars) as well as the DELPHI Monte Carlo (black histogram) and our \texttt{Whizard} simulation (green histogram) are shown.
The peak at $x_\gamma\sim 0.8$ corresponds to the process $e^+e^-\to\gamma Z\to\gamma\nu\bar\nu$ with an on-shell $Z$.
The red dotted histogram corresponds to the signal photon spectrum for the process $e^+e^-\to XX\bar X\bar X\gamma$, for $g=3$ and $M=20$~GeV.

In order to derive exclusion limits, we compare our signal + background energy spectrum with data by constructing a binned $\chi^2$ function defined as
\begin{equation}
\chi^2=\sum_{i=1}^{20}\frac{\left[N_{\text{data},i}-(N_{\text{SM},i}+N_{\text{NP},i})\right]^2}{\delta N_i^2}\simeq\sum_{i=1}^{20}\frac{N_{\text{NP},i}^2}{\delta N_i^2}\,,
\end{equation}
where $N_{\text{SM},i}$ corresponds to the number of simulated SM background events in the bin $i$, which have been taken to be equal to the number of data events $N_{\text{data},\,i}$, as first approximation. $N_{\text{NP},\,i}\equiv N_{\text{NP},\,i}(M,\,m_X,\,g)$ is the number of signal events and $\delta N_i$ is the total uncertainty on the number of events. 
For each point $(M,\,m_X,\,g)$ of the parameter space we construct the reduced chi squared $\chi^2_\text{dof}\equiv\chi^2/20$, the point is considered excluded at $95\%$~CL if $\chi^2_\text{dof}>1.57$.

\begin{figure}[t!]
\centering
\includegraphics[height=5.8cm]{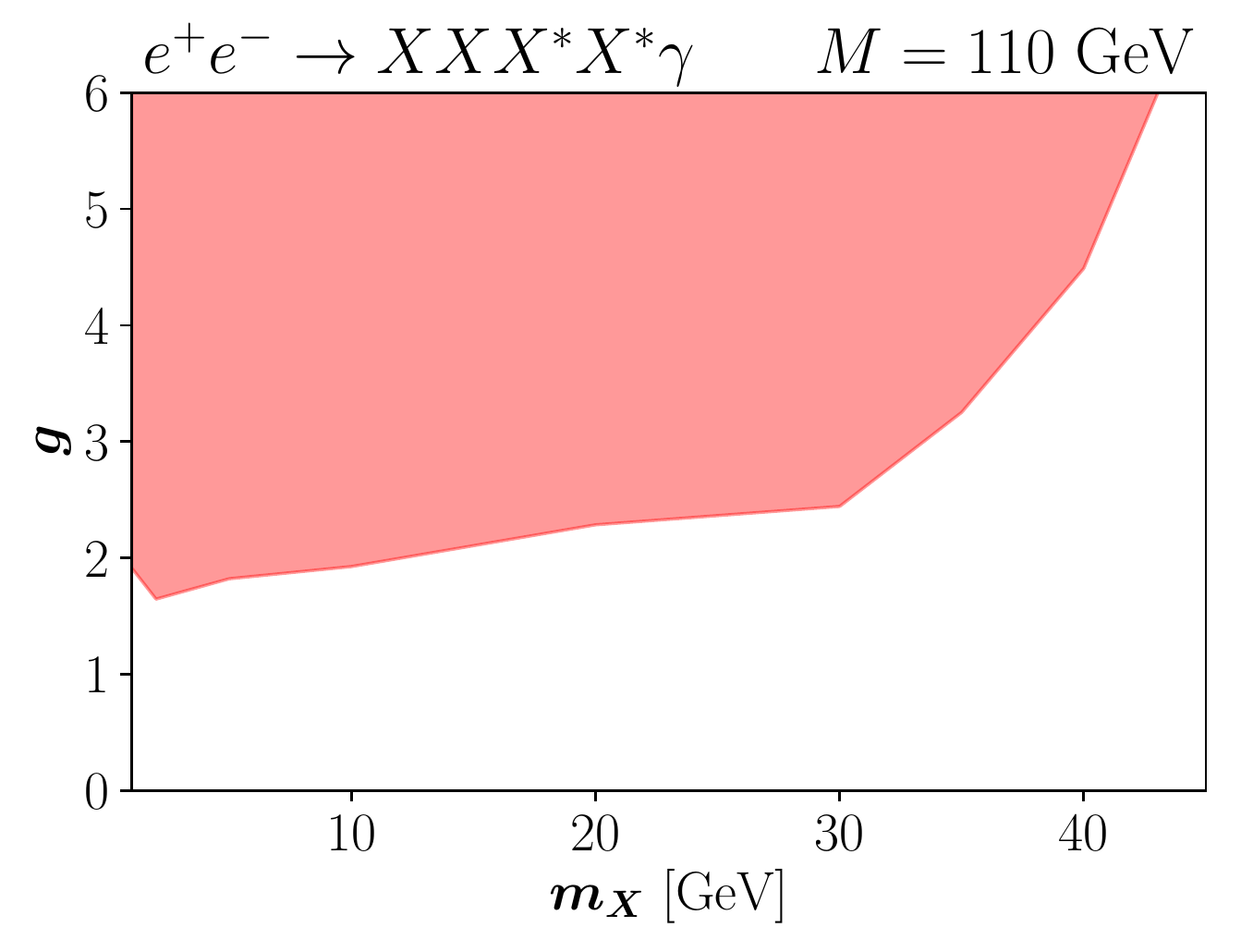}
\includegraphics[height=5.8cm]{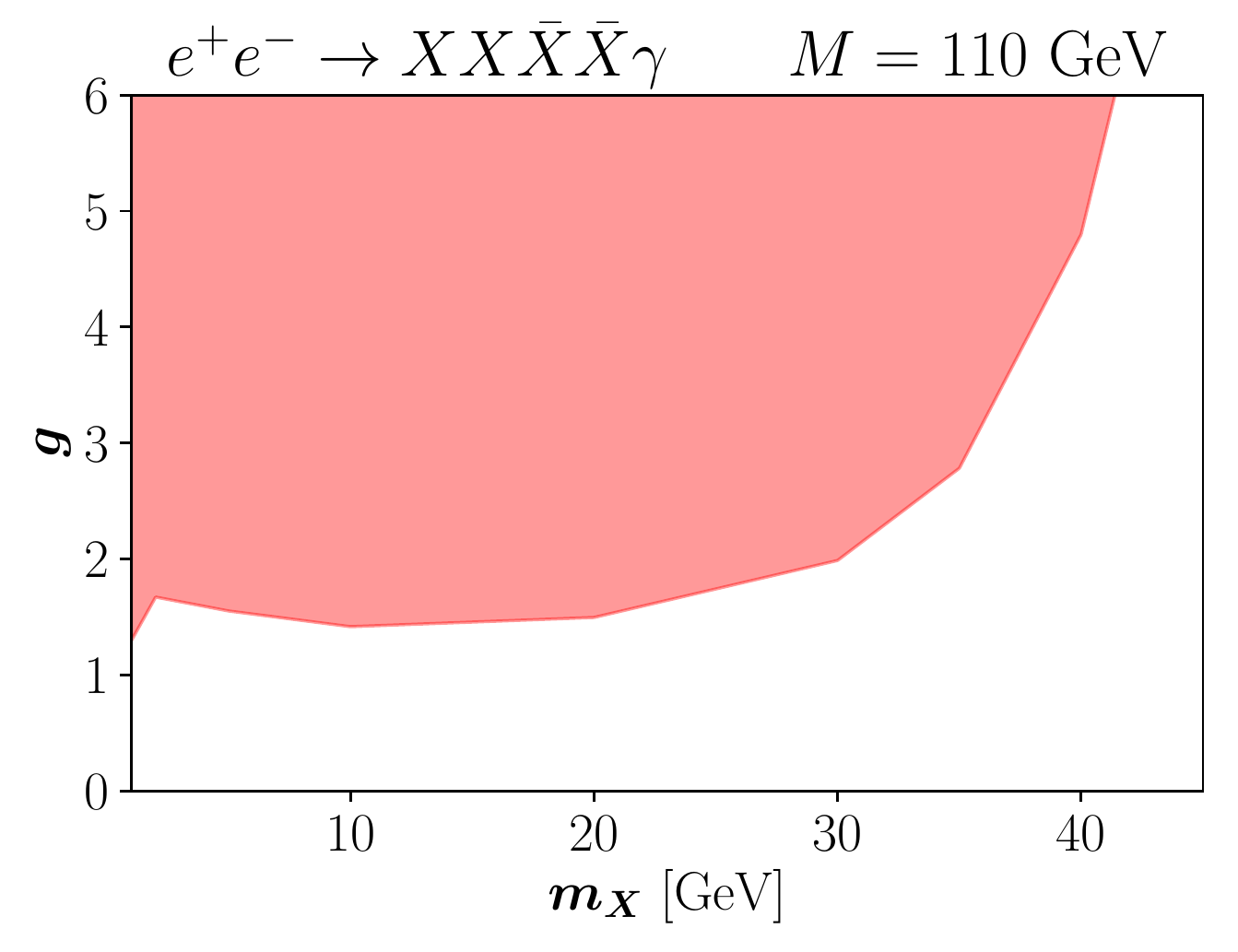}
\caption{The red shaded area represents the excluded region in the $(m_X,\,g)$ plane for scalar DM (left) and fermionic DM (right).}
\label{gxxxx}
\end{figure}

\begin{figure}[t!]
\centering
\includegraphics[height=5.2cm]{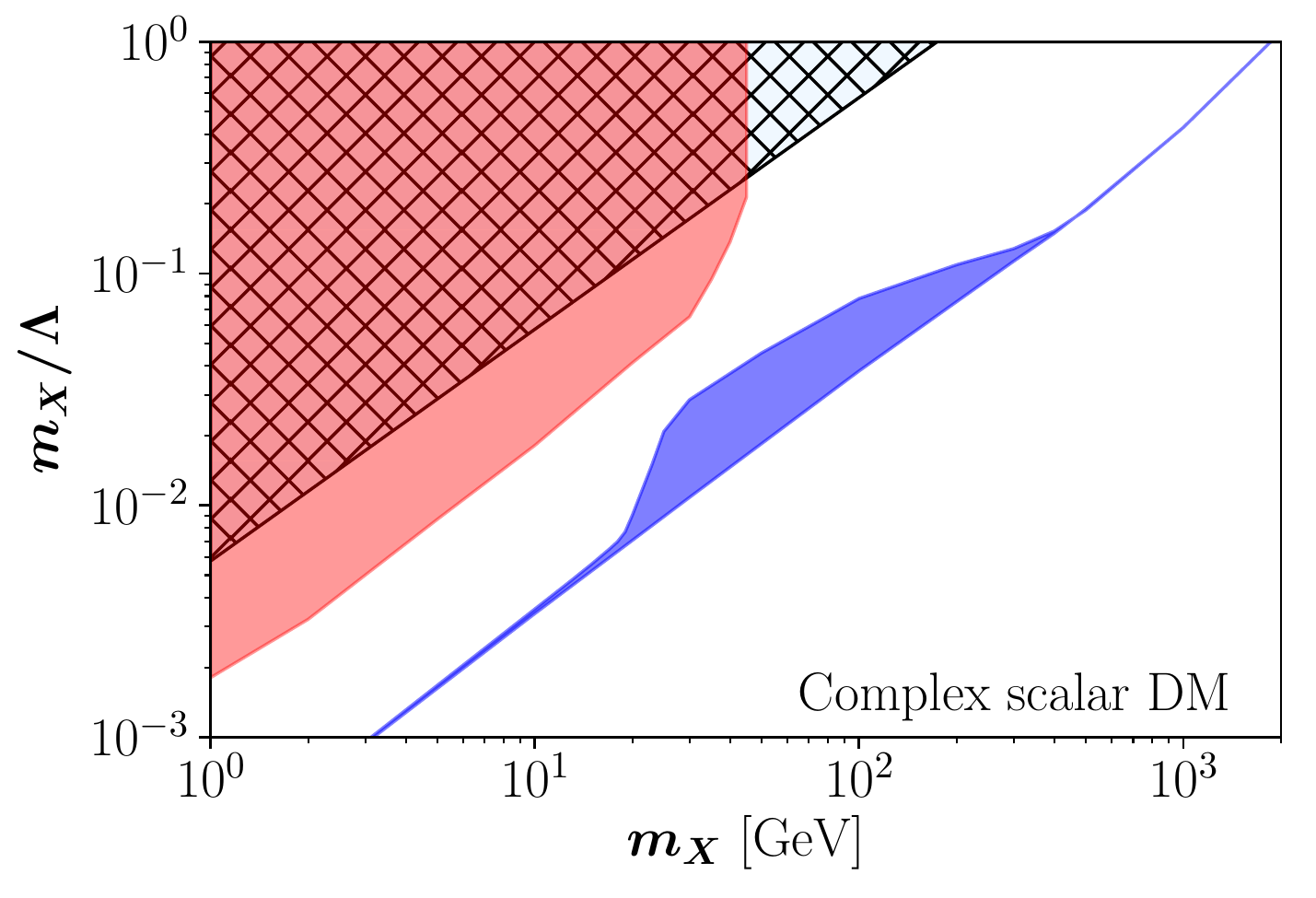}
\includegraphics[height=5.2cm]{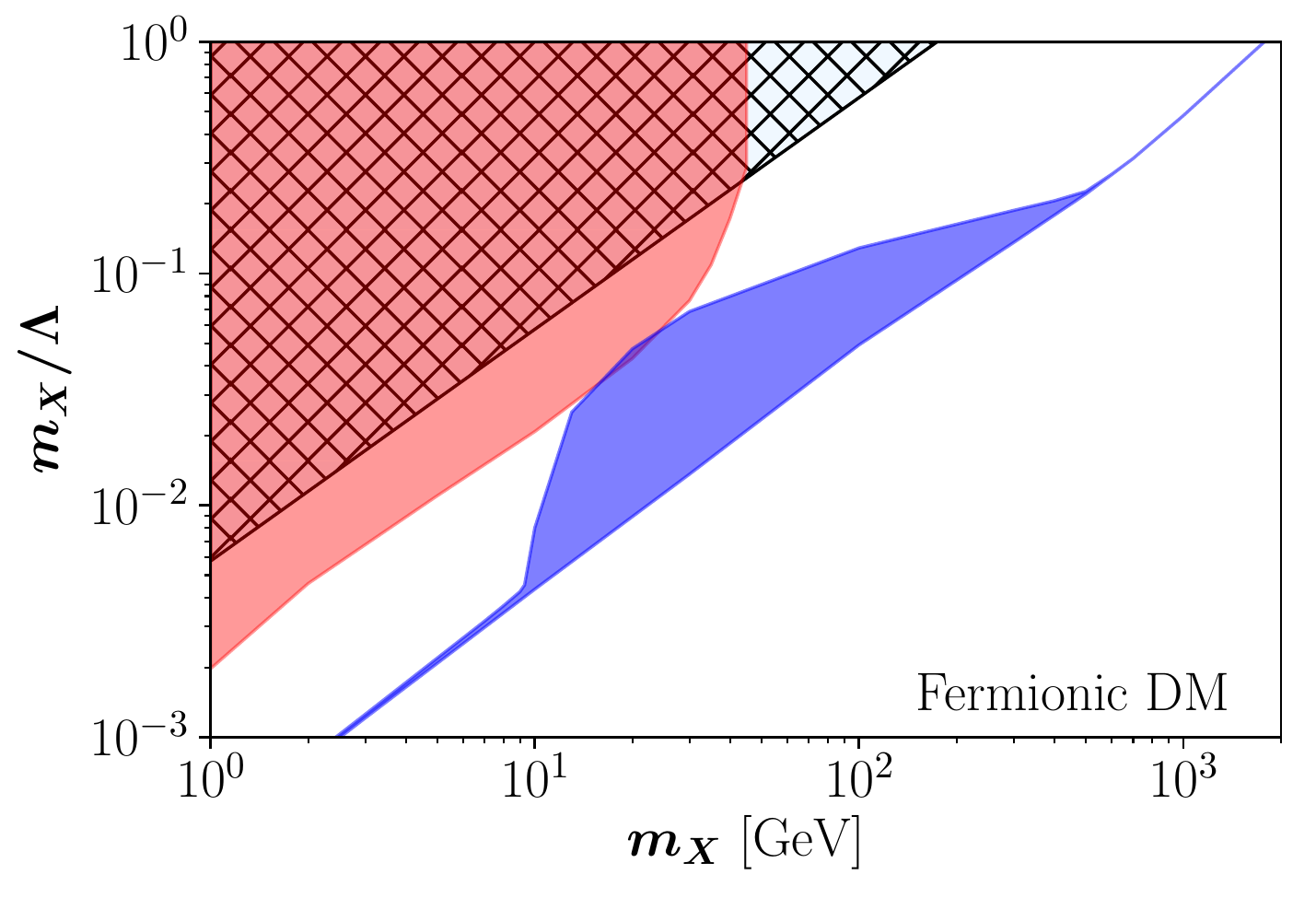}
\caption{The red shaded area represents the excluded region in the $(m_X,\,m_X/\Lambda)$ plane for scalar DM (left) and fermionic DM (right).}
\label{okmola}
\end{figure}

We determine the excluded regions in the $(m_X,\,g)$ plane for $m_X<50$~GeV, as shown in Fig.~\ref{gxxxx}. Using eq.~\eqref{boundnext} these results have been translated into ($\Lambda ,\,m_X/\Lambda$) plane, as shown in  Fig.~\ref{okmola}, and compared with the allowed parameter space for the considered sharing scenario.


\section{Scenarios with Light Mediators}
\label{sec:new_operator}

If we take some of the mediators $\phi_1$, $\phi_2$, $\psi$ to be lighter than $\sim 100$~GeV, the effective operator~\eqref{eq:op6} used in the analysis of ref.~\cite{Bernal:2016gfn} might not be valid after EWSp processes freeze out at around the same scale. As we have seen in Section~\ref{sec:di-jet}, the di-jet observable requires $\phi_1$ to be heavier than a few TeV while in Section~\ref{sec:LEP2}, the LEP bound also requires the $SU(2)$ doublet Dirac fermion $\psi$ to be above $\sim 100$~GeV~\cite{Achard:2001qw}. Hence we are left with the possibility that only the singlet scalar $\phi_2$ can have mass $m_{\phi_2} \lesssim 100$~GeV. 

If $\phi_2$ is short-lived (compared to the time needed by the asymmetry transfer to be completed), one can consider the same effective operators as in eq.~\eqref{eq:op6} for the asymmetry sharing as in ref.~\cite{Bernal:2016gfn} but with $\phi_2$ in the effective vertex `resolved', i.e. the cross sections have to be calculated with the $\phi_2$ propagator. 
Since the result is expected to be similar, it will not be considered here but instead we will focus on the case where $\phi_2$ is relatively long-lived. In this case, one can consider the following new sharing operator
\begin{equation}
{\cal O}' = \frac{1}{\Lambda^3} \phi_2^* \, {\cal O}^{(6)},
\label{eq:new_operator}
\end{equation}
where ${\cal O}^{(6)}$ represents one of the operators in eqs.~\eqref{eq:after_EW_op1}-\eqref{eq:after_EW_op4}. The matching of our UV model to the operator above is then
\begin{equation}
\frac{1}{\Lambda^3} = \frac{\lambda\, \eta\, \zeta}{m_{\phi_1}^2\,m_\psi}.
\label{match_new_operator}
\end{equation}

From eq.~\eqref{eq:new_operator}, the relevant scattering that we need to consider is $2 \leftrightarrow 3$: $\phi_2 \bar f \leftrightarrow f f f$ and $\bar f \bar f \leftrightarrow \phi_2^* f f$, where $f$ represents the four SM fermions in ${\cal O}^{(6)}$. Notice that the result of this sharing operator will be independent of whether $X$ is scalar or fermion. 
For this description to be valid, we have to make sure that $\phi_2$ particles only decay after the asymmetry sharing is completed. We further require that the decay rate for $\phi_2 \to X X$ ($m_X < m_{\phi_2}/2$) dominates over that of $\phi_2 \to ffff$ such that the asymmetry in $\phi_2$ will be transferred dominantly to $X$ through the decays $\phi_2 \to X X$ and the conjugate process. Without this assumption, we will not have an asymmetric DM. 
Hence we impose two requirements:
($i$) $\Gamma(\phi_2 \to X X) = H(T_d)$ for $T_d < T_{\rm fo}$ where $T_d$ and $T_{\rm fo}$ are respectively the decay temperature and asymmetry sharing freeze-out temperature; 
($ii$) $\Gamma(\phi_2 \to X X) \gg \Gamma(\phi_2 \to ffff)$. 
The requirement ($i$) gives the condition\footnote{The partial decay width of $\phi_2 \to X X$ is 
\begin{equation*}
\Gamma(\phi_2\to XX) = \frac{\mu^2}{8\pi} m_{\phi_2}
\sqrt{1 - \frac{4\,m_X^2}{m_{\phi_2}^2}} 
\, \left( 1 - \frac{2\,m_X^2}{m_{\phi_2}^2}\right),
\end{equation*}
where the last factor in the bracket is for the case of fermion $X$ and absent for scalar $X$. In estimating the bound, we have
dropped $m_X$ for simplicity.
}
\begin{equation}
\label{eq:bound1}
\mu \lesssim 2\times 10^{-8}
\sqrt{\frac{100\,{\rm GeV}}{m_{\phi_2}}}
\left(\frac{g_\star}{106.75}\right)^{1/4}
\left(\frac{3}{z_{\rm fo}}\right) .
\end{equation}
From the requirement ($ii$), we have\footnote{Here we estimate the partial decay width for $\phi_2 \to ffff$ to be 
\begin{equation*}
\Gamma(\phi_2 \to ffff) \approx \frac{1}{2^{12}\pi^5}
\frac{m_{\phi_2}^7}{\Lambda^6}.
\end{equation*}
}
\begin{equation}\label{eq:bound2}
\frac{m_{\phi_2}}{\Lambda} \ll 1 \times 10^{-2}
\left(\frac{\mu}{10^{-8}}\right)^{1/3}.
\end{equation}
In our numerical calculation, we take the above suppression to be 1/4 which corresponds to suppression of $\sim 10^{-4}$ in the branching ratio for $\phi_2 \to ffff$.

From eq.~\eqref{eq:new_operator} together with eqs.~\eqref{eq:after_EW_op1}-\eqref{eq:after_EW_op4}, the baryon and lepton number of $\phi_2$ is fixed to be $B_{\phi_2} = L_{\phi_2} = 1$. In the following, we will use the quantity $Y_x \equiv n_x/s$ where $n_x$ denotes the number density of $x$ and 
$s = (2 \pi^2/45)\,g_\star\,T^3$ the total entropic density with $g_\star$ the relativistic degrees of freedom. 
The Boltzmann equation to describe the evolution of $\phi_2$ asymmetry, $Y_{\Delta \phi_2} \equiv Y_{\phi_2} - Y_{\bar \phi_2} $, is given by
\begin{eqnarray}\label{BEafter}
sHz \frac{Y_{\Delta \phi_2}}{dz} 
 & = &  - \gamma
\left(\frac{Y_{\Delta \phi_2}}{ Y_{\phi_2}^{\rm eq}}
-\frac{1}{c_0\,Y_0}\Big[
c_B\, Y_{\Delta B} + c_L\, Y_{\Delta L}
-(c_B+c_L)\,Y_{\Delta \phi_2}
\Big]\right)\, ,
\end{eqnarray}
where $z \equiv m_{\phi_2}/T$, $H = 1.66 \sqrt{g_\star}\,T^2/M_{\rm Pl}$ the Hubble expansion rate, and $Y_0 \equiv 15/(8\pi^2 g_\star)$ with $M_{\rm Pl} = 1.22 \times 10^{19}$~GeV while 
$Y_{\Delta B}$ and $Y_{\Delta L}$ denote respectively the total baryon and lepton number asymmetries (sum of that of the SM and the dark sector).
We further fix $\{c_0,\,c_B,\,c_L\}=\{228,\,67,\,30\}$\footnote{For further details on these coefficients, please refer to ref.~\cite{Bernal:2016gfn}.} while the statistical function $\zeta_{\phi_2}$ is given by
\begin{equation}
\zeta_{\phi_2} = \frac{6}{\pi^{2}}\int_{z}^{\infty}dx\,x\,
\sqrt{x^{2}-z^{2}}\,\frac{e^{x}}{\left(e^{x}\pm 1\right)^{2}}\,.
\label{eq:zeta}
\end{equation}
The thermally averaged reaction density $\gamma \equiv\gamma_{\phi_2 \bar f\to fff}+\gamma_{\bar f\bar f\to \phi_2^* f f}$ denotes the sum over distinct scattering processes resulted from operator~\eqref{eq:operator} with ${\cal O}^{(6)}$ given by any of the operators~\eqref{eq:after_EW_op1}-\eqref{eq:after_EW_op4}.
The corresponding reduced cross section is presented in Appendix~\ref{app:reduced_cross_sections}.
Here we will solve with operator~\eqref{eq:after_EW_op1} assuming for simplicity a coupling only to first generation SM fermions. We take into account the gauge multiplicity $c_G = 3!\times 2 = 12$ and the possible distinct scattering processes where there is an additional factor of 3 for $\gamma_{\phi_2 \bar f\to fff}$ and additional factor of 6 for $\gamma_{\bar f\bar f\to \phi_2^* f f}$. The result for other operators~\eqref{eq:after_EW_op2}-\eqref{eq:after_EW_op4} can be obtained by rescaling with appropriate multiplicative factors.

\begin{figure}[t!]
\centering
\includegraphics[height=5.3cm]{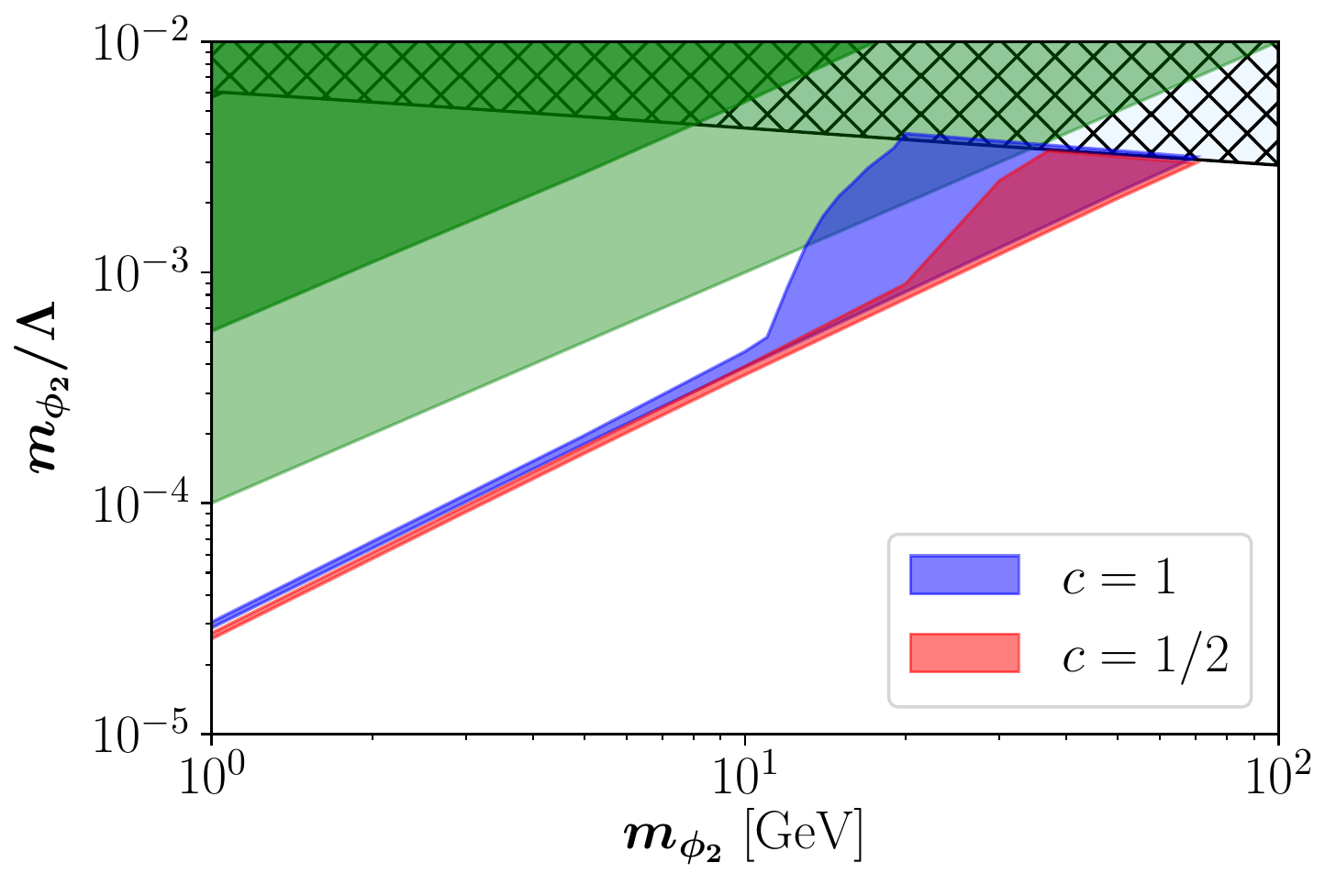}
\caption{Regions of the $(m_{\phi_2},\,\Lambda)$ plane characterized by the fact that the measured baryon asymmetry of the Universe and the DM relic abundance can be reproduced simultaneously, for $c=1$ (blue region) and $c=1/2$ (red region).
The upper hatched region corresponds to the bound in eq.~\eqref{eq:bound1} together with eq.~\eqref{eq:bound2}, such that our consideration of an asymmetric $\phi_2$ giving rise to an asymmetric DM $X$ remains valid. 
The light green region is excluded by di-jet bound in the \emph{simplest} model. 
The dark green region is excluded by mono-photon measurements at LEP 
in the \emph{next-to-simplest} model.
}
\label{mp2}
\end{figure}

Notice that besides the requirement $m_X < m_{\phi_2}/2$, $m_X$ is otherwise a free parameter. We parameterize $m_X = c\, m_{\phi}/2$ where $c < 1$. In this case, the total baryon asymmetry $Y_{\Delta B}$ carried by both the SM and the DM $X$ is
\begin{equation}
Y_{\Delta B} 
= \left(\frac{5.4\,m_n}{c \, m_{\phi_2}} + 1 \right) Y_{B_{\rm SM}}^0 ,
\end{equation}
where we used the center value of eq.~\eqref{eq:BAU} for $Y_{B_{\rm SM}}^0$.
Additionally, $Y_{\Delta L}$ is not fixed and we will vary it from $-\frac{51}{28} Y_{\Delta B}$
to 0 where the former is the equilibrium value due to the EWSp processes at $T \sim 100$~GeV 
while the latter takes into account the possibility of a mechanism which erases the lepton asymmetry. 
From the decays $\phi_2 \to X X$ and the conjugate process, we have $Y_{\Delta X} = 2\,Y_{\Delta \phi_2}$.
For our calculation, we fix 
$c = 1$ (limit case) and $c=1/2$,
and as the initial condition, either all 
the asymmetry resides in dark sector or in the SM sector i.e. either $Y_{\Delta \phi_2}(T_i) = Y_{\Delta B}$ 
or $Y_{\Delta \phi_2}(T_i) = 0$ where we take $T_i = 132$~GeV~\cite{D'Onofrio:2014kta}.
Given a $m_X$ (and hence $m_{\phi_2}$), we can determine $\Lambda$ by solving eq.~\eqref{BEafter} with the constraint that the final distribution of asymmetries matches the observed values. 
Our result is presented in Fig.~\ref{mp2}, for $c=1$ (blue region) and $c=1/2$ (red region).

\subsection{Phenomenological Constraints}
\label{sec:mphi2_lt_100}
In the \emph{simplest} version of the UV model where couplings and masses are taken to be equal (besides $m_{\phi_2}$) we have that
\begin{equation}
\frac{1}{\Lambda^3}=\frac{g^3}{M^3}.
\end{equation}
If we interpret the di-jet bound in terms of the \emph{simplest} version of the model eq.~\eqref{bounddijet} with $g=0.5$, we obtain
\begin{equation}
\Lambda > 10~{\rm TeV.}
\end{equation}
In this case, we can exclude part of the viable parameter space as shown in the light green region of Fig.~\ref{mp2}. 

If we consider the \emph{next-to-simplest} model for which $\zeta=\eta=g$, 
the matching condition becomes
\begin{equation}
\frac{1}{\Lambda^3}=\frac{\lambda\,g^2}{m_{\phi_1}^2\,m_\psi}.
\end{equation}
Additionally, if we interpret the di-jet bound from eq.~\eqref{bounddijet} with $\lambda = 0.5$ and $m_\psi = 110$~GeV, we have
\begin{equation}\label{boundnextnew}
\Lambda \gtrsim \frac{1.9}{g^{2/3}}~{\rm TeV.}
\end{equation}

\begin{figure}[t!]
\centering
\includegraphics[height=5.5cm]{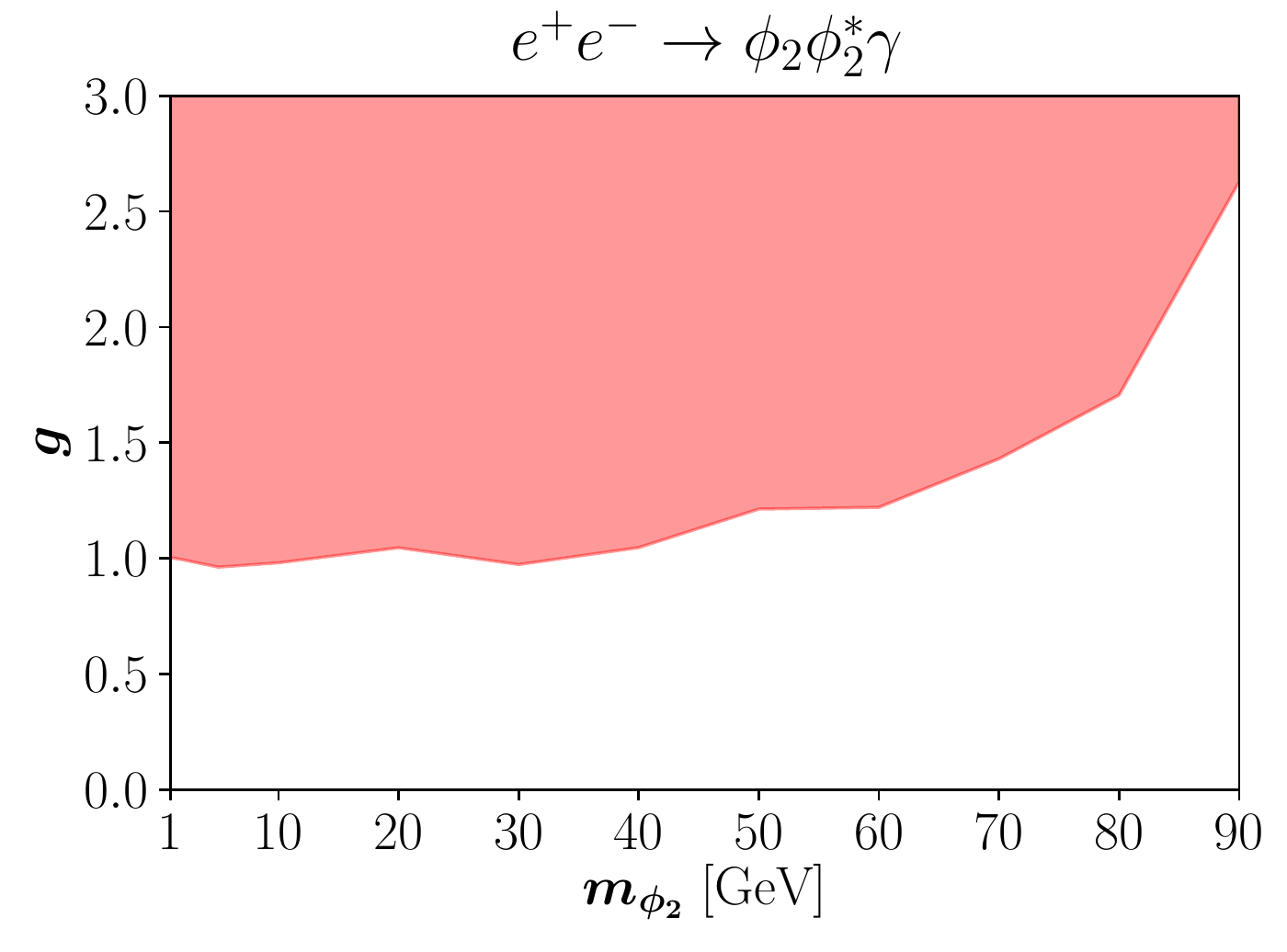}
\caption{Exclusion limits at $95\%$ CL for the process $e^+e^-\to\phi_2\phi_2^*\gamma$.
}
\label{phiphigamma}
\end{figure}

As in Section~\ref{sec:di-jet}, the di-jet production cross section by itself does not allow to extract a bound on $\Lambda$. In order to do this we consider another observable that allows to constrain $g$ independently. This observable is the photon production in association with MET at LEP. In this scenario, the model contributes to this signature through the process $e^+ e^- \to \phi_2 \phi_2^* \gamma$. 
We carry out the same analysis as laid out in Section~\ref{sec:LEP2}. 
We determine the excluded region in the $(m_{\phi_2},\,g)$ plane as shown in Fig.~\ref{phiphigamma}.
Using eq.~\eqref{boundnextnew} these results have been translated into ($m_{\phi_2},\,m_{\phi_2}/\Lambda$) plane, as shown in the dark green region of Fig.~\ref{mp2}. 
In this case, none of the viable parameter space of the new asymmetry sharing scenario is being excluded.

\section{Conclusions}
\label{sec:conclude}

In this work, we have considered the collider phenomenology of scenarios where the SM and DM particles share a common asymmetry through some effective operators in the early Universe.
Here we closely followed ref.~\cite{Bernal:2016gfn} and considered the case where DM is a singlet under the SM gauge interactions but carries nonzero baryon and/or lepton numbers. In this case, the DM can be
asymmetric just like the SM baryons, and the asymmetries could be \emph{shared}. We assumed
then the DM to be \emph{maximally} asymmetric, and either a complex scalar or a Dirac fermion.
The DM mass spans the range between few GeV and $\sim 100$~TeV. The connection between
the dark and the visible sectors is described by effective operators in the context of an
EFT, and it is separated in two different regimes depending on whether the transfer of the
asymmetries is effective before or after the EWSp processes freeze out.
The leading operators consisting of only the SM fields come in a limited number: one dim-5 operator with $B-L$ charge and four dim-6 operators with $B$ charge.

We have analyzed the LHC and LEP constraints on these effective operators by considering only events in which the partonic center of mass energy is below the mass scale of these operators such that the EFT description remains valid. We have showed that a portion of parameter space can still be excluded. Then, we proceeded to consider some representative UV complete models of these operators and showed that depending on the assumptions on the couplings and mass of the mediators of the models, the constraints derived can change, in some cases much stringent than the other.

\acknowledgments
NB was partially supported by the Spanish MINECO under Grant FPA2017-84543-P and by the European Union's Horizon 2020 research and innovation programme under the Marie Skłodowska-Curie grant agreements 674896 and 690575; and by the Universidad Antonio Nariño grants 2017239 and 2018204.
The work of AT was partially supported by Coordenação de Aperfeicoamento de Pessoal de N\'ivel Superior (CAPES). AT would like to thank ICTP-SAIFR and IFT-UNESP for hospitality.
CSF was supported by the São Paulo Research Foundation (FAPESP) under grants 2012/10995-7 \& 2013/13689-7 and is currently supported by the Brazilian National Council for Scientific and Technological Development (CNPq) grant 420612/2017-3.
AT and CSF thank the IIP, Natal for the organization of the workshop `LHC Chapter II: The Run for New Physic', where part of the progress was made.
We thank Andy Buckley, Andr\'e Lessa, Kentaro Mawatari and Darren Price for useful discussions.
In addition to the software packages cited above, this research made use of IPython~\cite{Perez:2007emg}, Matplotlib~\cite{Hunter:2007ouj}, and SciPy~\cite{SciPy}.

\appendix

\section{Thermally Averaged Reaction Densities}
\label{app:reduced_cross_sections}

Assuming Maxwell-Boltzmann phase space distribution, the thermally averaged reaction density for 2-to-$N$ scattering can be written as a function of the center of mass energy $\sqrt{s}$ and the temperature $T$ of the thermal bath as follows
\begin{equation}
\gamma_{ab\to ij...}  =  
\frac{T}{64\pi^{4}}\int_{s_{\rm min}}^\infty ds\sqrt{s}\,\hat{\sigma}_{ab\to ij...}\left(s\right)
{\cal K}_{1}\left(\frac{\sqrt{s}}{T}\right),\label{eq:reaction_density_sim}
\end{equation}
where $s_{\rm min}= \max\left[\left(m_{a}+m_{b}\right)^{2},
\left(m_{i}+m_{j}+...\right)^{2}\right]$, 
${\cal K}_1$ is the modified Bessel function of the second kind of order 1 
and the dimensionless reduced cross section is
\begin{equation}
\hat{\sigma}_{ab\to ij...}\left(s\right)  \equiv  2s\left[\beta\left(\frac{m_{a}^{2}}{s},\frac{m_{b}^{2}}{s}\right)\right]^{2}\sigma_{ab\to ij...}\left(s\right),\label{eq:reduced_cross_section}
\end{equation}
with $\sigma_{ab\to ij...}(s)$ the standard cross section and
\begin{equation}
\beta\left(v,w\right)\equiv \sqrt{\left(1-v-w\right)^2-4vw}.
\end{equation}

In term of more convenient parameter $z \equiv m/T$, 
$x \equiv s/m^2$ and $x_{\rm min} \equiv s_{\rm min}/m^2$ 
with $m$ is some convenient mass scale, eq.~\eqref{eq:reaction_density_sim} becomes
\begin{equation}
\gamma_{ab\to ij...}  =
\frac{m^4}{64\pi^{4} z}\int_{x_{\rm min}}^\infty ds\sqrt{x}\,\hat{\sigma}_{ab\to ij...}\left(x\right)
{\cal K}_{1}\left(\sqrt{x}z\right).
\end{equation}

The relevant reduced cross sections for Section~\ref{sec:new_operator} with $m = m_{\phi_2}$ are 
\begin{eqnarray}
\hat\sigma_{\phi_2\bar f \to fff}(x) &=& 
c_G \frac{m_{\phi_2}^6}{\Lambda^6}
\beta\left(x^{-1},0\right)
\frac{1}{2^7\pi^3} \frac{x^2 (x-1)}{12}, \\
\hat\sigma_{\bar f\bar f \to \phi_2^* ff}(x) &=&
c_G \frac{m_{\phi_2}^6}{\Lambda^6}
\frac{1}{2^7\pi^3} \frac{x^3}{2} ,
\end{eqnarray}
where $c_G$ is the gauge multiplicity and
we have kept only the mass of $\phi_2$
while all other fermions $f$ are taken to be massless.

\bibliographystyle{JHEP}
\bibliography{biblio}

\end{document}